\documentclass[12pt]{article}
\usepackage{graphics,bm,hyperref}
\rmfamily
\begin{document}
\begin{sloppypar}

\title{Magnetic Interactions in Transition-Metal Oxides}

\author{I. V. Solovyev$^{1,2}$ \\
$^1$Institute for Solid State Physics, University of Tokyo,\\
Kashiwanoha 5-1-5, Kashiwa 277-8531, Japan \\
$^2$ PRESTO, JST
}

\maketitle

\newpage

\begin{abstract}
The correct understanding of the nature and dynamics
of interatomic magnetic interactions in solids
is fundamentally important. In addition to that it allows to address and solve
many practical questions such as stability of equilibrium
magnetic structures, designing of magnetic phase diagrams,
the low-temperature spin dynamics, etc.
The magnetic transition temperature is also related with the behavior of interatomic
magnetic interactions.
One of the most interesting classes of magnetic compounds, which exhibits the rich
variety of the above-mentioned properties in the transition-metal oxides.
There is no doubts that all these properties are related with details of the
electronic structure.
In the spin-density-functional theory (SDFT),
underlying many modern first-principles electronic structure methods, there is a certain number of
fundamental theorems, which in principles provides a solid theoretical basis for the analysis of the
interatomic magnetic interactions. One of them is the magnetic force theorem, which connects
the total energy change with the change of single-particle energies obtained from solution of
the Kohn-Sham equations for the ground state.
The basic problem is that in practical implementations SDFT is always supplemented
with additional approximations, such as local-spin-density approximation (LSDA),
LSDA + Hubbard $U$, etc.,
which are not always adequate for the transition-metal oxides.
Therefore, there is not perfect methods,
and the electronic structure we typically have to deal with is always approximate. The main purpose of
this article, is to show how this, sometimes
very limited information about the electronic structure extracted
from the conventional calculations can be used for the solution of several practical
questions, accumulated in the field of magnetism of the transition-metal oxides.
This point will be illustrated for colossal-magnetoresistive manganites,
double perovskites, and magnetic pyrochlores.
We will review both successes and traps existing in the first-principle electronic
structure calculations, and make connections with the models which capture the basic physics of the
considered compounds. Particularly, we will show what kind of problems can be solved by
adding the Hubbard $U$ term on the top of the LSDA description.
It is by no means a panacea from all existing problems of
LSDA, and one should
clearly distinguish the cases when $U$ is indeed indispensable, play a minor role,
or may even lead to the systematic error.
\end{abstract}

\large

\newpage

\section{Introduction}
\label{Sec.Introduction}

  The first-principles electronic structure calculations play a very important role
in the exploration of magnetism. They have been successfully applied for various types
of metallic compounds. The transition-metal oxides (TMO), however, take a very special
place in this classification and typically regarded as a counter-example where the
first-principles calculations either experience serious
difficulties or simply do not work. Such an extreme point of view has of course a very
serious background because most of modern computational techniques are based
on the spin-density-functional theory (SDFT), designed for the ground state,
which is typically supplemented with the
local-spin density approximation (LSDA) for the exchange-correlation interactions.
The latter is based on the homogeneous electron gas theory, and therefore is very
different from the localized-orbital limit, which was originally adopted for the
description of TMO~\cite{Goodenough,AndersonHasegawa,Kanamori,Anderson}.

  Therefore, traditionally there was a big gap in the understanding of interatomic
magnetic interactions in TMO basing on these model
arguments
and the first-principles electronic structure calculations~\cite{Oguchi,Terakura},
which appeared only in the middle of 1980s and were regarded as very
challenging at that time.

  Since then, two different standpoints have certainly became closer.
It is true that due to complexity
of the problem of exchange and correlations, even now there is no perfect
computational methods for the
transition-metal oxides. However, it also becomes increasingly clear that
the model analysis of the problem should use results of first-principles
electronic structure calculations, at least as a starting point.
In many cases, a puzzling behavior attributed to the fanciful correlation
effects can be naturally explained by details of the realistic electronic
structure.

  The purpose of this article is to make a link between general formulation
of SDFT and models of interatomic magnetic interactions for TMO. We will
try to show that
even in the present form the electronic structure calculations can play a very
important role in the understanding of magnetic properties of various TMO,
despite many limitations inherent to LSDA and some of its refinements.

  In Sec.~\ref{Sec.Theory} we will summarize main results of SDFT and formulate
the magnetic force theorem which presents the basis for the analysis of
magnetic interactions in solids. The connection of this theorem
with some canonical models for interatomic magnetic interactions
will be illustrated in Sec.~\ref{Sec.models}.
In Sec.~\ref{Sec.MnO} we will present a critical analysis of interatomic
magnetic interactions in MnO basing on several available methods of
electronic structure calculations. In Sec.~\ref{Sec.applications} we will
consider some practical problems related with
the colossal-magnetoresistive (CMR)
manganites, double perovskites Sr$_2$Fe$M$O$_6$ ($M$$=$ Mo and Re), and magnetic
pyrochlores $A_2$Mo$_2$O$_7$ ($A$$=$ Y, Gd, and Nd).

  The insulating character of many
TMO presents one of the most interesting and controversial
problems. Particularly, it is well known that LSDA frequently underestimates or even
fails to reproduce the energy gap, which is formally the excited state property.
Is it possible that even in this case, it can provide a physically meaningful
description for the parameters of the ground state, such as the interatomic
magnetic interactions? In this context,
we will consider the role played by the on-site Coulomb interaction $U$
on the top of the LSDA electronic structure
and argue that
one should clearly distinguish the behavior of \textit{band insulators}, which is
governed by the double exchange (DE) physics~\cite{AndersonHasegawa,deGennes},
and in principle can be accounted
for by itinerant models, in the spirit of LSDA, from the behavior of \textit{Mott insulators},
where the Coulomb $U$ is indeed indispensable in order to suppress (in this case) spurious
DE interactions and unveil the completely different physical behavior governed by the
superexchange (SE) interactions~\cite{Kanamori,Anderson,KugelKhomskii}.
A brief summary and perspectives will
be outlined in Sec.~\ref{Sec.Summary}.

\section{Theory of magnetic interactions in solids}
\label{Sec.Theory}

\subsection{Spin-Density-Functional Theory}
\label{SubSec.SDFT}

  The modern way to approach the problem of interatomic magnetic interactions in
solids is based on the SDFT~\cite{SDFT},
which states that the magnetic ground state of the $N$-electron system can be
obtained by minimizing the Hohenberg-Kohn total energy functional
\begin{equation}
E [ {\bf m} ] = T_0[ {\bf m} ] + E_{\rm XC}[ {\bf m} ]
\label{eqn:Em}
\end{equation}
($T_0[ {\bf m} ]$ being the kinetic energy of a non-interaction electron system
and $E_{\rm XC}[ {\bf m} ]$ being the exchange-correlation energy)
with respect to the spin-magnetization density ${\bf m}({\bf r})$.\footnote{For the sake
of simplicity, we have dropped in Eq.~(\ref{eqn:Em}) the
electron density, $n({\bf r})$, and all terms which depend only on $n({\bf r})$,
though this dependence is implied as well as the minimization of
$E$ with respect to $n({\bf r})$.}

  The direct implementation of the SDFT
is hampered by the fact
that the functional dependencies $T_0[ {\bf m} ]$ and $E_{\rm XC}[ {\bf m} ]$
are generally unknown. In the case of $T_0[ {\bf m} ]$,
the problem is resolved by introducing an auxiliary system of one-electron orbitals
$\{ \psi_i({\bf r}) \}$, and requesting the kinetic energy
(in Rydberg units),
$$
T_0[ {\bf m} ] = \sum_{i=1}^{N} \int d {\bf r} \psi_i^\dagger({\bf r})
\left( - \nabla^2 \right) \psi_i({\bf r}),
$$
the spin-magnetization density,
\begin{equation}
{\bf m}({\bf r}) = \sum_{i=1}^{N} \psi_i^\dagger({\bf r}) \bm{\sigma} \psi_i({\bf r})
\label{eqn:sdensity}
\end{equation}
($\bm{\sigma}$ being the vector of Pauli matrices), and the total energy
to coincide with the same parameters of the real many-electron system
in the
ground state. Then, the minimization of $E [ {\bf m} ]$ with respect to ${\bf m}({\bf r})$
is equivalent to the self-consistent solution
of single-particle Kohn-Sham (KS) equations for $\{ \psi_i({\bf r}) \}$:
\begin{equation}
\left[ - \nabla^2 + \bm{\sigma} \cdot {\bf B}({\bf r}) \right] \psi_i({\bf r}) =
\varepsilon_i \psi_i({\bf r}),
\label{eqn:KS}
\end{equation}
where the effective magnetic field is given by
$$
{\bf B}({\bf r}) = \frac{\delta}{\delta {\bf m}({\bf r})} E_{\rm XC}[ {\bf m} ].
$$

  The exchange-correlation energy functional, $E_{\rm XC}[ {\bf m} ]$, is typically taken
in an approximate form. Several possible approximations along this line are listed below.
\begin{itemize}
\item
The local-spin-density approximation. In this case, the explicit dependence
of $E_{\rm XC}$ on ${\bf m}({\bf r})$ and the electron density $n({\bf r})$,
\begin{equation}
E_{\rm XC}[ n, {\bf m} ] = \int d {\bf r} n({\bf r}) \varepsilon_{\rm XC}
\left[ n({\bf r}), |{\bf m}({\bf r})| \right],
\label{eqn:LSDA}
\end{equation}
is borrowed from the theory of homogeneous electron gas. Conceptually,
LSDA is similar to the Stoner theory of band magnetism~\cite{Stoner}.
Due to the rotational invariance
of the homogeneous electron gas, $E_{\rm XC}$ depends only on the absolute
value of ${\bf m}({\bf r})$.
\item
The local-density approximation plus Hubbard $U$ (LDA$+$$U$) approach~\cite{AZA,PRB94,PRL98}.
This is a semi-empirical approach, the main idea of which is to cure some shortcomings
of the LSDA description for the localized electron states by replacing corresponding part of
$E_{\rm XC}[ n, {\bf m} ]$ in LSDA by the energy of on-site Coulomb interactions,
in an analogy with
the multi-orbital Hubbard model. The latter is typically
taken in the mean-field Hartree-Fock form:
$$
E_{\rm XC}[\widehat{n}^{\bm{\tau}}] = \sum_{\{ \gamma \}}
\left( U_{\gamma_1 \gamma_3 \gamma_2 \gamma_4} - U_{\gamma_1 \gamma_3 \gamma_4 \gamma_2} \right)
n_{\gamma_1 \gamma_2}^{\bm{\tau}} n_{\gamma_3 \gamma_4}^{\bm{\tau}},
$$
where $U_{\gamma_1 \gamma_3 \gamma_2 \gamma_4}$$\equiv$$\langle \gamma_1 \gamma_3 | \frac{1}{r_{12}} |
\gamma_2 \gamma_4 \rangle$ are the matrix elements of the on-site Coulomb interactions,
which are assumed to be renormalized from the bare atomic values by interactions with other
(itinerant) electrons and by correlation effects in solids. The electron and spin-magnetization
densities for the localized states are represented by corresponding elements of the
density matrix $\widehat{n}^{\bm{\tau}}$$=$$\| n_{\gamma_1 \gamma_2}^{\bm{\tau}} \|$
in the basis of atomic-like orbitals $\{ \gamma \}$ at the site ${\bm{\tau}}$.
Because of this construction, the LDA$+$$U$ approach is basis-dependent and
typically implemented in the linear muffin-tin orbital method~\cite{LMTO}.
\item
The optimized effective potential (OEP) method~\cite{OEP,Kotani}
is an attempt of exact numerical solution
of the KS problem, which does not rely on the local-spin-density approximation
for $E_{\rm XC}[ n, {\bf m} ]$.
In this case, the eigenfunctions $\{ \psi_i({\bf r}) \}$ and eigenvalues
$\{ \varepsilon_i \}$ obtained from the KS equations (\ref{eqn:KS})
with some trial potential
are used as an input for total energy calculations beyond the
homogeneous electron gas limit.\footnote{
Of course, practical implementations of the OEP scheme require some approximations
for the exchange-correlation energy. Typically, it is either Hartree-Fock or the
the random-phase approximation, underlying
the $GW$ method~\cite{GW_review}).}
The parameters of such potential are requested to minimize
the total energy.
\end{itemize}

\subsection{Magnetic Force Theorem}
\label{SubSec.MFT}

   The basic idea behind analysis of interatomic magnetic interactions in solids
is to evaluate the total energy change
$\Delta E$$=$$E[\widehat{R}_{\bm{\theta}}{\bf e}_{\rm GS}]$$-$$E[{\bf e}_{\rm GS}]$ caused
by small non-uniform rotations of the spin-magnetization density near the ground
state,
where
${\bf e}_{\rm GS}({\bf r})$$=$${\bf m}_{\rm GS}({\bf r})/|{\bf m}_{\rm GS}({\bf r})|$
is the direction of the
spin-magnetization in the ground state and
$\widehat{R}_{\bm{\theta}}$ is the three-dimensional rotation
by the (small) angle $\bm{\theta}({\bf r})$, which depends on the position ${\bf r}$
in the real space:
$$
\widehat{R}_{\bm{\theta}({\bf r})}{\bf e}_{\rm GS}({\bf r}) = {\bf e}_{\rm GS}({\bf r}) +
[ \bm{\theta}({\bf r}) \times {\bf e}_{\rm GS}({\bf r})]
- \frac{1}{2} \bm{\theta}^2({\bf r}) {\bf e}_{\rm GS}({\bf r}).
$$
Thus,
characterizes the local stability of the magnetic ground state with respect to
non-uniform rotations of the spin-magnetization density.

  Very generally, the problem can be solved by minimizing the constrained total
energy functional:
$$
E_{\bf h}[{\bf m}] = T_0[{\bf m}] + E_{\rm XC}[ {\bf m} ] -
\int d {\bf r} {\bf h}({\bf r}) \cdot \left[ {\bf e}({\bf r}) -
\widehat{R}_{\bm{\theta}({\bf r})} {\bf e}_{\rm GS}({\bf r}) \right],
$$
where the constraining field ${\bf h}({\bf r})$ plays the role of
Lagrange multipliers and
enforces the requested distribution of the spin-magnetization density
in the real space.

  The magnetic force theorem states in this respect: in the second
order of $\bm{\theta}({\bf r})$ the total energy change is solely determined
by the change of the KS single-particle energies~\cite{PRB98},
\begin{equation}
\Delta E = \sum_{i=1}^N \left(
\varepsilon_i \left[  \widehat{R}_{\bm{\theta}}{\bf B}, \widehat{R}_{\bm{\theta}}{\bf m}_{\rm GS} \right] -
\varepsilon_i \left[ {\bf B},{\bf m}_{\rm GS} \right] \right) + O \left( \bm{\theta}^2 \right).
\label{eqn:LFT}
\end{equation}
The eigenvalues
$\varepsilon_i \left[  \widehat{R}_{\bm{\theta}}{\bf B}, \widehat{R}_{\bm{\theta}}{\bf m}_{\rm GS} \right]$,
corresponding to the rotated effective field $\widehat{R}_{\bm{\theta}}{\bf B}$ and
the ground state spin-magnetization density
$\widehat{R}_{\bm{\theta}}{\bf m}_{\rm GS}$,
can be expressed in terms of expectation values of the KS Hamiltonian,
$$
\varepsilon_i \left[  \widehat{R}_{\bm{\theta}}{\bf B}, \widehat{R}_{\bm{\theta}}{\bf m}_{\rm GS} \right] =
\int d {\bf r} \psi_i^\dagger [\widehat{R}_{\bm{\theta}}{\bf m}_{\rm GS}]
\left( - \nabla^2 + \bm{\sigma} \cdot \widehat{R}_{\bm{\theta}}{\bf B} \right)
\psi_i [\widehat{R}_{\bm{\theta}}{\bf m}_{\rm GS}],
$$
with $\{ \psi_i [\widehat{R}_{\bm{\theta}}{\bf m}_{\rm GS}] \}$
yielding
$\widehat{R}_{\bm{\theta}}{\bf m}_{\rm GS}$ after substitution into Eq.~(\ref{eqn:sdensity}).

  The theorem can be reformulated in a different way: the effect of the longitudinal change
of ${\bf m}({\bf r})$ on $T_0[{\bf m}]$ and $E_{\rm XC}[ {\bf m} ]$
caused by the self-consistent solution of the KS equations (\ref{eqn:KS})
with the external magnetic field ${\bf h}({\bf r})$ is cancelled out
in the second order of $\bm{\theta}({\bf r})$.

  The theorem can be proven rather generally provided that the exchange-correlation
energy functional obeys the following condition~\cite{PRB98}:
\begin{equation}
E_{\rm XC}[ {\bf m} ] = E_{\rm XC}[ \widehat{R}_{\bm{\theta}} {\bf m} ].
\label{eqn:gauge}
\end{equation}
In LSDA, it immediately follows from Eq.~(\ref{eqn:LSDA}).
More generally, Eq.~(\ref{eqn:gauge}) can be regarded as the fundamental
gauge-symmetry constraint, which should be superimposed on the admissible form of the
exchange-correlation energy functionals~\cite{VignaleRasolt}.\footnote{
Since ${\bf m}({\bf r})$ is given by Eq.~(\ref{eqn:sdensity}), the three-dimensional rotation
${\bf m}({\bf r})$$\rightarrow$$\widehat{R}_{\bm{\theta}({\bf r})}{\bf m}({\bf r})$ is
equivalent to the unitary transformation of the KS orbitals
$\psi_i({\bf r})$$\rightarrow$$\widehat{U}_S \left[ \bm{\theta}({\bf r}) \right] \psi_i({\bf r})$,
where $\widehat{U}_S \left[ \bm{\theta}({\bf r}) \right]$$=$$\exp \left[ \frac{i}{2}
\bm{\sigma} \cdot \bm{\theta}({\bf r}) \right]$ is the $2$$\times$$2$ rotation matrix in the
spin subspace. Then, the property (\ref{eqn:gauge}) can be easily proven for many other
functionals, for example the ones based on the Hartree-Fock approximation and underlying the
rotationally invariant LDA$+$$U$~\cite{PRL98} and
OEP~\cite{Kotani} methods.}

  The magnetic force theorem has very important consequences:
\begin{itemize}
\item
Generally, the KS eigenvalues $\{ \varepsilon_i \}$ have no physical meaning and cannot
be compared with the true single-particle excitations (that is typically the case for
many spectroscopic applications).
In this respect,
the magnetic excitations present a pleasant exception, thanks to the magnetic force theorem.
\item
In principle, the knowledge of the effective KS potential alone is sufficient to
calculate the total energy difference. It allows to get rid of
heavy total energy calculations
(particularly, for the OEP method) without any loss of the accuracy.
\end{itemize}

\subsection{Practical Implementations of the Magnetic Force Theorem}
\label{SubSec.practice}

  In most cases, practical applications of the magnetic force theorem are based on
relaxed constraint conditions in comparison with the ones given by
Eq.~(\ref{eqn:LFT}). Namely, the second condition requesting the KS eigenvalues
to correspond the rotated spin magnetization density is typically dropped
and $\varepsilon_i \left[  \widehat{R}_{\bm{\theta}}{\bf B}, \widehat{R}_{\bm{\theta}}{\bf m}_{\rm GS} \right]$
is replaced by $\varepsilon_i \left[  \widehat{R}_{\bm{\theta}}{\bf B} \right]$ obtained from
the KS equations (\ref{eqn:KS}) with the rotated field
$\widehat{R}_{\bm{\theta}}{\bf B}$.
This considerably facilitates the calculations, though at the cost of a systematic error,
which was recently discussed by Bruno~\cite{Bruno}.\footnote{
Since ${\bf B}({\bf r})$$=$$\frac{\delta}{\delta {\bf m}({\bf r})} E_{\rm XC}[ {\bf m} ]$
and $E_{\rm XC}[ {\bf m} ]$ satisfies Eq.~(\ref{eqn:gauge}), the rotation of the
spin-magnetization density
${\bf m}_{\rm GS}({\bf r})$$\rightarrow$$\widehat{R}_{\bm{\theta}({\bf r})}{\bf m}_{\rm GS}({\bf r})$
results in similar rotation of the effective field
${\bf B}({\bf r})$$\rightarrow$$\widehat{R}_{\bm{\theta}({\bf r})}{\bf B}({\bf r})$.
Therefore, at the input of the KS equations (\ref{eqn:KS}) the direction of the
spin-magnetization is consistent with the direction of the effective field. However,
the direction of the new magnetization obtained after the first iteration can be canted off
the initial distribution prescribed by the matrix $\widehat{R}_{\bm{\theta}}$.
This is precisely the source of the error, and strictly speaking the magnetic force theorem
cannot be proven for this relaxed constrained condition. In order to correct this error,
Bruno~\cite{Bruno} explicitly considered the constraining field
${\bf h}({\bf r})$, which can be estimated using the response-type arguments.
This field does not explicitly contribute to the expression (\ref{eqn:LFT})
for the total energy change. However, it does modify the KS orbitals
$\{ \psi_i \}$, which should be used in order to evaluate
$\varepsilon_i \left[  \widehat{R}_{\bm{\theta}}{\bf B}, \widehat{R}_{\bm{\theta}}{\bf m}_{\rm GS} \right]$.
Although
the analysis undertaken by Bruno
is certainly valid and the finding is important, we believe that there
is some confusion with the terminology, because the magnetic force theorem itself is correct,
as long as it is formulated in the form of Eq.~(\ref{eqn:LFT})~\cite{PRB98}. The main
criticism by Bruno~\cite{Bruno} is devoted to the \textit{practical implementations}
of this theorem.}
We will return to this problem in Sec.~\ref{SubSubSec.OLa}
where will give some estimates of this error and discuss some physical
implications which may be related with the additional
rotation of the spin-magnetization density for the transition-metal oxides. In this section
we will review some practical schemes of calculations,
which ignore these effects. Basically, they are two.

  The first one is based on the perturbation theory expansion for the Green
function~\cite{Liechtenstein}
$$
G({\bf r},{\bf r}',\varepsilon) = \sum_{i} \frac{\psi_i({\bf r}) \psi_i^\dagger({\bf r}')}
{\varepsilon - \varepsilon_i + i\delta}
$$
and its projections
$$
G^{\uparrow,\downarrow}=\frac{1}{2} {\rm Tr}_S \left\{ (1 \mp \sigma_z ) G \right\}
$$
on the majority ($\uparrow$) and minority ($\downarrow$) spin states (${\rm Tr}_S$
being the trace over the spin indices). In the second order of
$\bm{\theta}({\bf r})$, the total energy change can be mapped onto the Heisenberg model
\begin{equation}
\Delta E = -\frac{1}{2} \int d {\bf r} \int d {\bf r}' J({\bf r},{\bf r}')
\left[ \widehat{R}_{\bm{\theta}({\bf r})}{\bf e}_{\rm GS}({\bf r}) \cdot
\widehat{R}_{\bm{\theta}({\bf r}')}{\bf e}_{\rm GS}({\bf r}')  -
{\bf e}_{\rm GS}({\bf r}) \cdot {\bf e}_{\rm GS}({\bf r}') \right].
\label{eqn:Heisenberg}
\end{equation}
The parameters of this model are given by~\cite{Liechtenstein}\footnote{
Here, the mapping onto the Heisenberg model is a general property,
whereas the from of Eq.~(\ref{eqn:Jij}) corresponds to the additional
approximation
$\varepsilon_i \left[  \widehat{R}_{\bm{\theta}}{\bf B}, \widehat{R}_{\bm{\theta}}{\bf m}_{\rm GS} \right]$$
\rightarrow$$\varepsilon_i \left[  \widehat{R}_{\bm{\theta}}{\bf B} \right]$
for the KS eigenvalues.}
\begin{equation}
J({\bf r},{\bf r}') = \frac{2}{\pi} {\rm Im} \int_{-\infty}^{\varepsilon_F}
d \varepsilon G^\uparrow ({\bf r},{\bf r}',\varepsilon) B({\bf r}')
G^\downarrow ({\bf r}',{\bf r},\varepsilon) B({\bf r}),
\label{eqn:Jij}
\end{equation}
where $\varepsilon_F$ is the Fermi energy corresponding to the highest occupied KS orbital.
All practical calculations along this line are typically performed on a discrete lattice,
assuming that all space is divided into atomic regions specified by the site indices
$\{ \bm{\tau} \}$, so that the angles $\bm{\theta}({\bf r})$ depend only on $\bm{\tau}$,
and neglecting the effects caused by rotations of the spin-magnetization density on the
intra-atomic scale.\footnote{
The non-collinear distribution of the spin-magnetization density on the intra-atomic
scale is an interesting and so far very
imperfectly understood phenomenon~\cite{intranc}.}
In the discrete version,
$G^{\uparrow,\downarrow} ({\bf r},{\bf r}',\varepsilon)$ is expanded
in the basis of atomic-like orbitals $\{ \gamma \}$ centered at different
atomic sites (for example, in the nearly-orthogonal LMTO
representation~\cite{LMTO}). The integral
over atomic regions around $\bm{\tau}$ and
$\bm{\tau}'$, $J_{\bm{\tau} \bm{\tau}'}$$=$$\int_{\bm{\tau}} d {\bf r} \int_{\bm{\tau}'} d {\bf r}'
J({\bf r},{\bf r}')$,
defines conventional parameters of interatomic magnetic interactions
in the real space.

  The main idea of the second approach is to calculate the energy of the collective
spin excitation corresponding to the frozen spin wave with the vector ${\bf q}$. The
method, which is called the frozen (or adiabatic) spin-wave approximation is designed
for the discrete lattice~\cite{Halilov}.
In this context, the adiabaticity means that the directions of the spin-magnetization
can be regarded as the ''slow'' variables, so that for each configuration specified by
$\widehat{R}_{\bm{\theta}}$ the ''fast'' electronic degrees of freedom
have enough time to follow
this directional distribution of the spin-magnetization density.
The rotation matrix $\widehat{R}_{\bm{\theta}}$ is requested to transform the
collinear distribution of the spin magnetic moments in the ground state,
${\bf m}_{\bm{\tau}}$$=$$\int_{\bm{\tau}} d {\bf r}~{\bf m}_{\rm GS} ({\bf r})$,
to the spin spiral: ${\bf m}_{\bm{\tau}}$$\rightarrow$$
(\cos {\bf q} \cdot \bm{\tau} \sin \theta_{\bm{\tau}},
\sin {\bf q} \cdot \bm{\tau} \sin \theta_{\bm{\tau}}, \cos \theta_{\bm{\tau}})
| {\bf m}_{\bm{\tau}} |$. If ${\bf m}_{\bm{\tau}}$$||$$z$, the rotation angles are
given by
$$
\bm{\theta}_{\bm{\tau}} = \left( -\sin {\bf q} \cdot \bm{\tau}, \cos {\bf q} \cdot \bm{\tau}, 0 \right)
\theta_{\bm{\tau}}
$$
($\theta_{\bm{\tau}}$ being the cone-angle of the spin wave).
The KS equations for the spin-spiral configuration
of the effective field $\widehat{R}_{\bm{\theta}} {\bf B}$
can be solved by
employing the generalized Bloch transformation~\cite{gBloch}. This gives
the total
energy change $\Delta E ( {\bf q},\bm{\theta})$ corresponding to the
''excited'' spin-spiral configuration with the vector
${\bf q}$. In the the second order of $\bm{\theta}$, this energy change
can be mapped onto the Heisenberg
model. The mapping provides parameters of magnetic interactions in the reciprocal space,
$J_{\bf q}$, which can be Fourier transformed to the real space.

\section{Relation with the Models of Magnetic Interactions}
\label{Sec.models}

  The application of Eq.~(\ref{eqn:Jij}) goes far beyond the standard electronic
structure calculations. It is rather general, and many canonical expressions for
magnetic interactions in solids can be derived starting with Eq.~(\ref{eqn:Jij}).
Here we would like to illustrate this idea by considering two model examples.

\subsection{Double Exchange and Superexchange Interactions in Half-Metallic
Ferromagnetic State}
\label{SubSec.modelDE}

\begin{figure}[h!]
\centering \noindent
\resizebox{7cm}{!}{\includegraphics{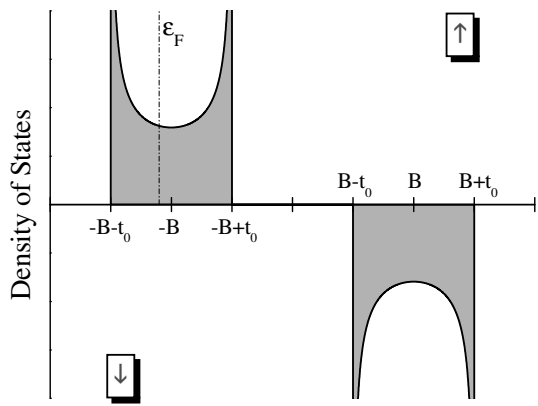}}
\caption{Schematic density of states for the half-metallic ferromagnetic chain.}
\label{fig.DOS}
\end{figure}

  Consider the ferromagnetic (FM) chain of atoms,
described by the Hamiltonian
$$
{\cal H}_{\tau \tau'} = -t_0 \delta_{\tau \pm 1,\tau'} + \sigma_z B,
$$
which is an analog of the KS Hamiltonian (\ref{eqn:KS}) on the discrete lattice,
where matrix elements of the kinetic energy $t_0$ (the transfer integrals)
are restricted by the nearest neighbors, and $B$ is
the effective
field polarizing the conduction electrons parallel to the $z$ axis.
The half-metallic behavior implies that $t_0$$<$$B$ and the electron
density $n$$<$$1$ (which corresponds to the
partial population of the $\uparrow$-spin band in Fig.~\ref{fig.DOS}).
The problem can be easily solved analytically, and by
expanding $G_{\tau \tau'}^{\downarrow}(\varepsilon)$ in Eq.~(\ref{eqn:Jij})
up to the second order of $t_0/B$
one can obtain the following contributions to the nearest-neighbor (NN)
exchange coupling~\cite{PRL99}:
\begin{equation}
J^D=\frac{t_0}{2\pi}\sin\pi n
\label{eqn:DE}
\end{equation}
and
\begin{equation}
J^S=-\frac{t_0^2}{2\pi B}(\pi n + \frac{1}{2}\sin 2 \pi n ).
\label{eqn:SE}
\end{equation}
$J^D$ is the FM double exchange interaction, which is proportional to $t_0$.
$J^S$ is the antiferromagnetic (AFM) superexchange interaction.
For the half-filled band at
$n$$=$$1$ the system is insulating. Then, $J^D$$=$$0$, while Eq.~(\ref{eqn:SE})
leads to the standard
expression
for the SE interaction at the half-filling:
$J^S$$=$$-$$\frac{t_0^2}{2B}$~\cite{Anderson}.

  More generally, $J^D$ and $J^S$ can be expressed through the moments of local density
of states, and $J^D$ is the measure of the kinetic energy of the fully spin-polarized
half-metallic states~\cite{Springer}. Originally, the concept of the double exchange
was introduced for the metallic phase of CMR manganites~\cite{deGennes}.
However, the phenomenon
appears to be more generic and one of the most interesting recent
suggestions was that the same mechanism can operate in the insulating state, provided that
the system is a \textit{band insulator}~\cite{Springer}. This substantially modifies the
original view on the problem and combine two seemingly orthogonal concepts, one of
which is the DE physics
and the other one is the insulating behavior of some CMR manganites.
We will return to this problem in Sec.~\ref{SubSec.DECMR}.

\subsection{Superexchange Interaction via Oxygen States}
\label{SubSec.modelSE}

\begin{figure}[h!]
\centering \noindent
\resizebox{7cm}{!}{\includegraphics{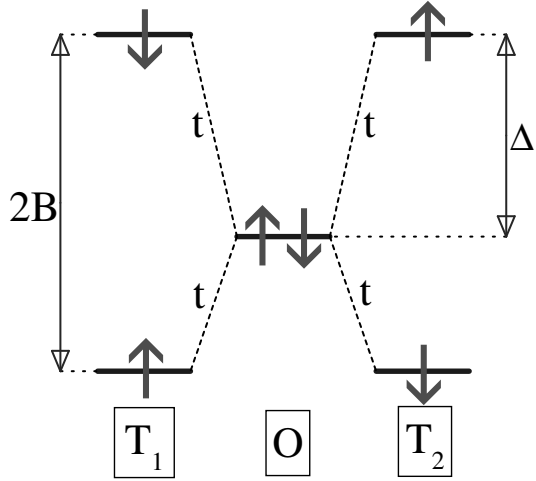}}
\caption{Positions of atomic levels illustrating the superexchange
interaction between transition-metal sites mediated by the oxygen states.}
\label{fig.SEmodel}
\end{figure}

  Consider the interaction between two magnetic (transition-metal) site,
which is mediated by the non-magnetic oxygen states (Fig.~\ref{fig.SEmodel},
the situation is rather common for the insulating transition-metal oxides).
It is assumed that the splitting between the $\uparrow$- and $\downarrow$-spin
states of the
transition metal sites is $2B$. $\Delta$ describes the
relative position of the oxygen states relative to the transition-metal
states (the so-called charge-transfer energy).
It is further assumed that the occupied states are T$_1$($\uparrow$), T$_2$($\downarrow$), and
O($\uparrow$,$\downarrow$), and the unoccupied states are T$_1$($\downarrow$)
and T$_2$($\uparrow$). $2B$ and $\Delta$
are the largest parameters in the problem, so that the transfer interactions $t$
between the transition-metal and oxygen sites can be treated as a perturbation
starting with the atomic limit.
Corresponding matrix elements of the Green function connecting the sites
T$_1$ and T$_2$ are given by
\begin{equation}
G^{\uparrow}_{12}(\varepsilon)=G^{\downarrow}_{21}(\varepsilon)=
t^2 ( \varepsilon+B)^{-1} ( \varepsilon-B+\Delta)^{-1}
( \varepsilon-B)^{-1}.
\label{eqn:GFSE}
\end{equation}
Then, using Eq.~(\ref{eqn:Jij})
one can obtain the well-known expression for the SE interactions mediated
by the oxygen states~\cite{SEoxygen}:\footnote{
Here
we have considered the AFM configuration of the sites T$_1$ and T$_2$
(Fig.~\ref{fig.SEmodel}). The assumption is not important and absolutely
the same expression for $J^S_{12}$ can be obtained by starting with the FM configuration.
This means that in the case of superexchange, the mapping of the total energy
change onto the Heisenberg model is universal. This is not a general rule,
and other types of magnetic interactions do depend on the state in which they
are calculated. This dependence is related with the change of electronic
structure which enters the expression for the magnetic interactions (\ref{eqn:Jij})
through the matrix elements of the Green function. As we will see in Sec.~\ref{SubSec.DECMR},
the form of the DE interactions strongly depends on the magnetic state and
this dependence plays a crucial role in the physics of CMR manganites.}
$$
J^S_{12}=-\frac{t^4}{\Delta^2} \left( \frac{1}{\Delta} + \frac{1}{2B} \right),
$$
which describes the shift of the poles of the Green function (\ref{eqn:GFSE})
of the occupied states, located at $\varepsilon$$=$$-$$B$ and
$\varepsilon$$=$$B$$-$$\Delta$, due to the interaction with the unoccupied states
located at $\varepsilon$$=$$B$.
\\

  Considered examples of DE and SE interactions
show that Eq.~(\ref{eqn:Jij}) is rather universal and can be regarded
as the starting point for the general analysis of interatomic magnetic interactions in solids.
The applications can be very wide and cover, for example, the theory of
RKKY interactions~\cite{RKKY},
the effects of the interatomic Coulomb interactions on the SE coupling~\cite{PRB99},
etc.

\section{Electronic Structure of MnO from the viewpoint of Interatomic Magnetic Interactions}
\label{Sec.MnO}

  The manganese monoxide provides an excellent opportunity to test available methods of
electronic structure calculations for the transition-metal oxides. While numerous
spectroscopic techniques deal mainly with the excited state properties, which cannot be
compared directly with results of electronic structure calculations designed for the
ground state, the interatomic magnetic interactions are the ground-state properties
and, in principle, should be accounted for by these calculations.

  The interatomic
magnetic interactions in MnO are well studied experimentally. The magnetic
behavior of MnO can be described by the simple Heisenberg model
including first ($J_1$$=$$-$$4.8$ meV) and second ($J_2$$=$$-$$5.6$ meV) neighbor
interactions~\cite{MnO}.\footnote{According to the definition (\ref{eqn:Heisenberg})
of the Heisenberg model,
the experimental parameters have been multiplied by
$S^2$$=$$(5/2)^2$. We do not consider the trigonal distortion of the cubic
lattice caused by the exchange striction effects~\cite{MnO}.}
Since MnO is a wide-gap insulator, it is clear that both interactions
originate from the superexchange mechanism~\cite{Anderson,Oguchi}.

  One can easily design
the proper model for the electronic structure of MnO
using the following arguments~\cite{PRB98}.
Since the electronic configuration of the Mn atoms in MnO is close to
$3d^5_{\uparrow} 3d^0_{\downarrow}$, both from the viewpoint of the model
valence arguments and the electronic structure calculations,
the distribution of the spin-magnetization density near Mn sites
is nearly spherical.
Therefore, the only parameter of the effective magnetic field at the Mn sites
we need to worry about is $B$, which controls the splitting between
occupied $3d_{\uparrow}$ and unoccupied $3d_{\downarrow}$ states.
Another important parameter of the electronic structure is the
charge-transfer energy $\Delta$, which controls the relative position of the
$3d_{\downarrow}$ and the oxygen $2p$ states.
Then, using an analogy with the model for SE interactions
considered in Sec.~\ref{SubSec.modelSE}, one can try to parameterize the
effective KS potential in terms of
$B$ and $\Delta$ (basically in the spirit of LDA$+$$U$ for the
half-filled $3d$ shell), and to find these \textit{two} parameters
by fitting \textit{two} experimental
parameters of interatomic magnetic interactions, $J_1$ and $J_2$,
provided that the kinetic and non-magnetic parts of the effective KS Hamiltonian
are well described on the level of LDA.

\begin{figure}[h!]
\centering \noindent
\resizebox{10cm}{!}{\includegraphics{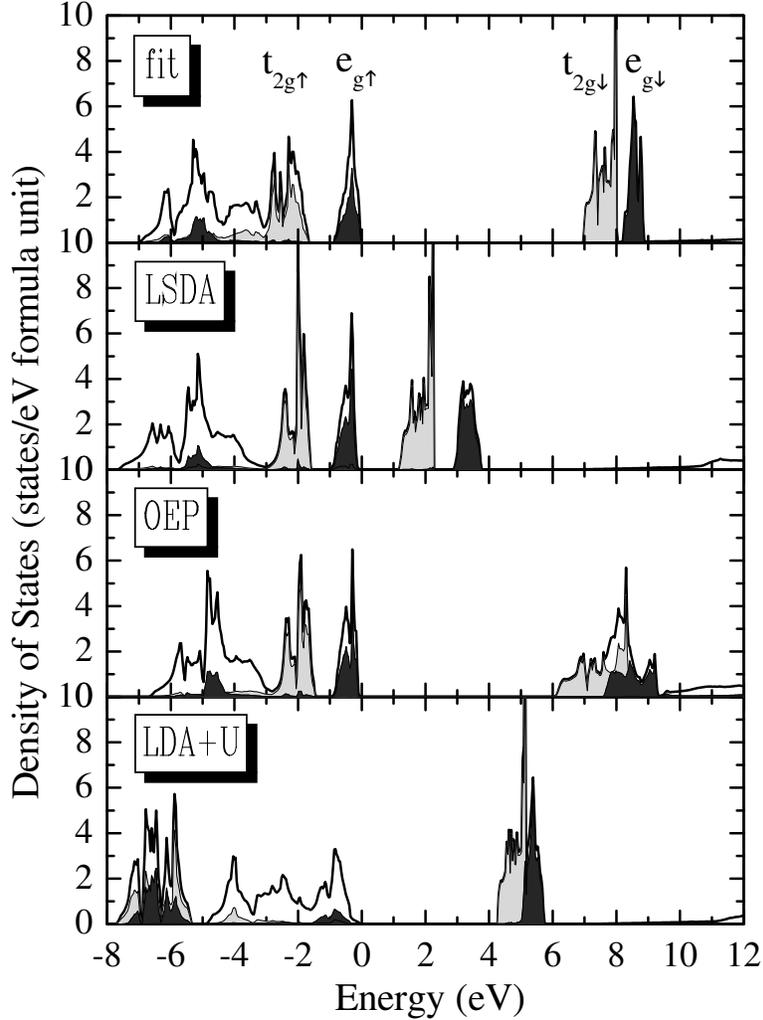}}
\caption{Electronic structure of MnO, which fits the experimental
parameters of magnetic interactions together with results of
first-principles calculations in LSDA, OEP, and LDA$+$$U$.}
\label{fig.MnO}
\end{figure}

  The fitting, which is very straightforward in the
nearly-orthogonal LMTO basis~\cite{LMTO,PRB98}, yields
the following parameters: $B$$\simeq$$5.3$ eV and $\Delta$$\simeq$$10.7$ eV.\footnote{
The on-site Coulomb repulsion $U$ for the $3d$ electrons
can be estimated from $2B$ by using the connection $2B$$=$$\frac{1}{5}(U+4J)\mu_{\rm Mn}$
which holds for the mean-field Hartree-Fock solution of the degenerate Hubbard model
($\mu_{\rm Mn}$$\simeq$$4.84$ $\mu_B$ being the local magnetic moment).
Then, using $J$$\simeq$$0.8$ eV for the intra-atomic exchange coupling,
which is not sensitive to the crystal environment in solid~\cite{AZA},
one obtains $U$$\simeq$$8$ eV.}
The corresponding density of states is shown in Fig.~\ref{fig.MnO}, together with
results of the LSDA, LDA$+$$U$, and OEP calculations.\footnote{
All OEP calculations discussed in this section have been performed using
the one-electron effective potentials obtained by T.~Kotani~\cite{Kotani}.}
Corresponding parameters of magnetic interactions are listed in Table~\ref{tab:MnO}.
\begin{table}[h!]
\caption{Parameters of magnetic interactions in MnO obtained in LSDA,
OEP and LDA$+$$U$ in comparison with experimental data~\protect\cite{MnO}.}
\label{tab:MnO}
\begin{center}
\begin{tabular}{lcc}\hline
                          & $J_1$ (meV)        & $J_2$ (meV)        \\\hline
LSDA                      & -13.2              & -23.5              \\
OEP                       & -8.9               & -12.0              \\
LDA$+U$                   & -5.0               & -13.2              \\
Expt.                     & -4.8               & -5.6     \\\hline
\end{tabular}
\end{center}
\end{table}
Unfortunately, none of the existing techniques can treat the problem of
interatomic magnetic interactions in MnO properly. Nevertheless, the analysis of
the electronic structures shown in Fig.~\ref{fig.MnO} allows us to elucidate the
basic problems of the existing methods which lead to substantial overestimation
of $J_1$ and $J_2$~\cite{PRB98}.
\begin{itemize}
\item
LSDA underestimates both $B$ and $\Delta$. In this sense, the overestimation of
interatomic magnetic interactions is directly related with the
underestimation of the band gap. However, the occupied part of the spectrum is
reproduced reasonably well by the LSDA calculations.
\item
LDA$+$$U$ underestimates $\Delta$. This is a very serious problem inherent not only to
LDA$+$$U$ but also to some subsequent methodological developments for the strongly correlated
systems like LDA+DMFT approach~\cite{DMFT}, which provides a solid basis
for the description of
Coulomb interactions
at the transition-metal sites,
but ignore other parameters of electronic structure, which may play an important role
for TMO.
\item
The relative position
of the $3d$ and $2p$ states is well reproduced by the OEP method.
However, the width of the unoccupied $3d$ band is
strongly overestimated. This might be related with the local form of the effective
one-electron potential in the OEP approach. As the result, the parameters
of interatomic magnetic interactions are strongly overestimated.
\end{itemize}

  Thus, the quantitative description of interatomic magnetic interactions
in TMO by the first-principle electronic structure
calculations is largely unresolved and challenging problem. The analysis of MnO
suggests that one possible direction along this line could be to borrow the
form of the one-electron potential from the LDA$+$$U$ approach and try to
optimize parameters of this potential by using the ideas of variational
OEP method~\cite{PRB98}. In any case, the development of the first-principle
techniques, which could address the problem
of electronic structure and magnetic interactions in TMO
on a more superior level is
certainly a very important direction for the future.

  However, even in the present form, the electronic structure calculations
may have a significant impact on our understanding of the physical properties of
TMO. In the next section we would like to
illustrate how this, sometimes rather limited information extracted from the
conventional electronic structure calculations can be used for the
analysis of many important questions accumulated in the field magnetism
of TMO.

\section{Applications for Transition-Metal Oxides}
\label{Sec.applications}

\subsection{Double Exchange Interactions in CMR Manganites}
\label{SubSec.DECMR}

\begin{figure}[h!]
\centering \noindent
\resizebox{7cm}{!}{\includegraphics{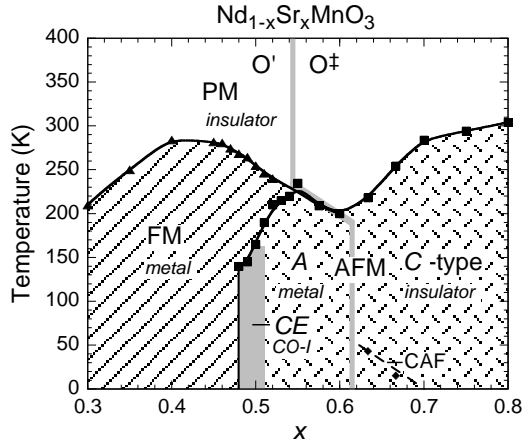}}
\caption{Magnetic phase diagram of Nd$_{1-x}$Sr$_x$MnO$_3$
(from Ref.~\protect\cite{NSMO}). Different phases are denoted as follows:
PM - paramagnetic, FM - ferromagnetic, and AFM - antiferromagnetic,
consisting of the FM layers (A), simple chains (C), and zigzag chains (CE).}
\label{fig.NSMO}
\end{figure}

  One of the most famous groups of oxide materials, which was in the focus of enormous
experimental and theoretical attention for many years, is perovskite
manganese oxides (or simply -- the manganites) with the chemical formulas
$R_{1-x}D_x$MnO$_3$ (cubic type),
$R_{1-x}D_{1+x}$MnO$_4$ (layered type),
and $R_{2-2x}D_{1+2x}$Mn$_2$O$_7$ (double-layered type;
$R$ and $D$ being trivalent rare-earth and divalent
alkaline-earth ions, respectively).
The manganites are known for the colossal magnetoresistance effect, that
is gigantic suppression of the resistivity by the external magnetic field.
Another important feature of these compounds is the rich magnetic
phase diagram (Fig.~\ref{fig.NSMO})
Typically, the CMR effect occurs at the boundary of either FM and PM, or
FM and AFM phases. In this sense, the understanding of the magnetic phase
diagram is directly related with the understanding of the CMR effect.
There is a large number of modern review articles devoted to these
compounds~\cite{Springer,Imada,GordonBreach,Dagotto},
which covers a lot of experimental information as well as the
theoretical views on the problem. In this section we would like to illustrate,
as we believe, the main idea behind the magnetic phase diagram
and the CMR effect itself from the viewpoint of electronic structure
calculations and the general theory of interatomic magnetic
interactions presented in Sec.~\ref{Sec.Theory}. We will concentrate on
the doping range $x$$\geq$$0.3$.\footnote{The magnetic phase diagram at small
$x$ is significantly affected by the Jahn-Teller effect. This is certainly
an interesting and not completely understood problem (see, e.g., Ref.~\cite{Springer}).
However, the physics is rather different from a more canonical
double exchange regime realized for $x$$\geq$$0.3$.}

  According to the formal valence arguments, the $3d$ states of Mn in
$R_{1-x}D_x$MnO$_3$ are filled by the $(4$$-$$x)$ electrons, which are
subjected to the strong Hund's rule coupling.
Therefore, the intra-atomic exchange splitting
between the $\uparrow$- and $\downarrow$-spin $3d$ states can be
regarded as the largest parameter
in the problem. The $3d$ states are further split by the cubic crystal-field into
the lower-lying $t_{2g}$ and higher-lying $e_g$ levels, yielding the formal
electronic configuration
$t_{2g \uparrow}^3 e_{g \uparrow}^{1-x} t_{2g \downarrow}^0 e_{g \downarrow}^0$.
The band dispersion, caused by the interatomic hopping interactions involving
$t_{2g}$ and $e_g$ orbitals is typically smaller than the intra-atomic
exchange and crystal-field splittings. The hoppings are mediated by the oxygen
$2p$ states, and in the undistorted cubic lattice are allowed only between
orbitals of the same, either $t_{2g}$ or $e_g$, symmetry.
The strengths of the Mn($e_g$)-O($2p$) and Mn($t_{2g}$)-O($2p$) interactions
are controlled by the Slater-Koster parameters, correspondingly $pd\sigma$ and $pd\pi$.
Since $|pd\sigma|$$>$$|pd\pi|$,
the $t_{2g}$ band
is typically narrower than the $e_g$ one~\cite{Priya}.

  Thus, the $t_{2g}$ states will be half-filled and according to Sec.~\ref{SubSec.modelDE}
contribute only to the AFM SE interactions.
The $e_g$ states are partially occupied and therefore will contribute
to both FM DE and AFM SE interactions, whose ratio will depend on $x$.

  What is the relevance of this picture to the magnetic phase diagram
of perovskite manganites? In order to get a very rough idea about this problem, let us simply
count the number of FM ($z_{\rm FM}$) and AFM ($z_{\rm AFM}$) bonds formed by the
Mn sites in each of the magnetic phases shown in Fig.~\ref{fig.NSMO}
(Table~\ref{tab:z}).
\begin{table}[h!]
\caption{Number of ferromagnetic ($z_{\rm FM}$) and antiferromagnetic ($z_{\rm AFM}$)
bonds for the main magnetic phases observed in Nd$_{1-x}$Sr$_x$MnO$_3$.}
\label{tab:z}
\begin{center}
\begin{tabular}{cccc}\hline
phase  & $x$ in phase diagram             & $z_{\rm FM}$       & $z_{\rm AFM}$      \\\hline
FM     & $0.3$$\leq$$x$$<$$0.5$    & 6                  & 0                  \\
CE     & $x$$\sim$$0.5$            & 2                  & 4                  \\
A      & $0.5$$\leq$$x$$\leq$$0.6$ & 4                  & 2                  \\
C      & $x$$>$$0.6$               & 2                  & 4                  \\
G      & $x$$\sim$$1$              & 0                  & 6                  \\\hline
\end{tabular}
\end{center}
\end{table}
The FM interactions prevail at smaller $x$.
However, further increase of $x$ leads to the gradual change of the character of
magnetic interactions towards the antiferromagnetism, which
is reflected in the increase of the
number of AFM bonds.
Therefore, it seems reasonable to link this phase diagram with some kind of competition
between FM and AFM interactions.

  In order to further proceed with this picture,
we need to address the following questions.
\begin{enumerate}
\item
Which mechanism stabilizes the anisotropic AFM A and C structures in the metallic regime?\footnote{
According to the experimental data, the A phase is metallic while the C phase is
an insulator due to the cation disorder
in the quasi-one-dimensional C-type AFM structure. Our point of view is that this disorder
is not directly related with the magnetic stability of the C phase, and as the first
approximation this phase can be treated in the conventional DE model which corresponds to the
metallic behavior.}
Note, that according to the canonical DE model~\cite{deGennes}, the metallic
antiferromagnetism is unstable with respect to a spin-canting.
\item
What is so special about the CE phase? Why does it exist only in the narrow window of
$x$ close to $x$$=$$0.5$ and perturbs the monotonous change of $z_{\rm FM}$ and
$z_{\rm AFM}$ shown in Table~\ref{tab:z}? Why is it insulator?
\item
Quantitative description of the doping-dependance of the magnetic phase diagram.
\end{enumerate}

\subsubsection{Basic Details of the Electronic Structure}
\label{SubSubSec.DECMRbasic}
\begin{figure}[h!]
\centering \noindent
\resizebox{10cm}{!}{\includegraphics{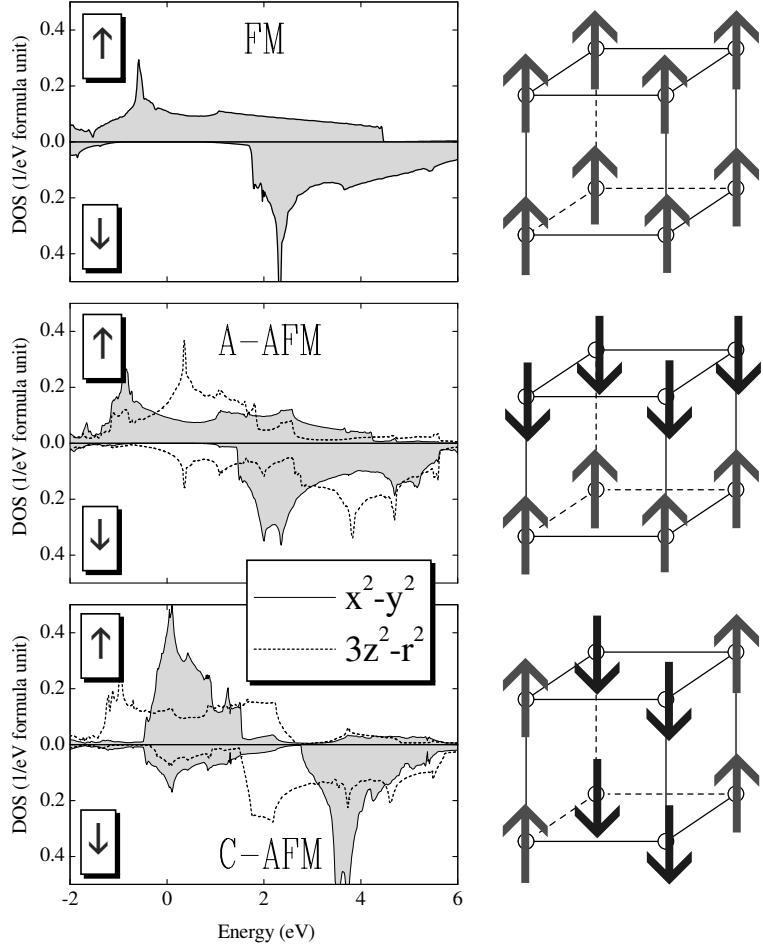}}
\caption{Distribution of the Mn($e_g$) states in the ferromagnetic, A-, and
C-type antiferromagnetic phases of La$_{1/2}$Ba$_{1/2}$MnO$_3$ in LSDA.
The Fermi level is at zero.}
\label{fig.BSMOdos}
\end{figure}

  The key factor factor, which allows to resolve the questions summarized in
parts 1 and 2 is the strong dependance of the electronic structure of the $e_g$
states on the magnetic structure~\cite{BrinkKhomskii,PRB01,CEpapers}.
The effect can be estimated in the LSDA by
considering the distribution of the $e_g$ states in the hypothetical virtual-crystal
alloy La$_{1/2}$Ba$_{1/2}$MnO$_3$ (Fig.~\ref{fig.BSMOdos}).
In this calculations, the crystal structure of La$_{1/2}$Ba$_{1/2}$MnO$_3$ was
rigidly fixed to be the cubic one. Hence, Fig.~\ref{fig.BSMOdos} shows the pure response of the
electronic structure of the $e_g$ states to the change of the magnetic structure.
Even in LSDA, which typically underestimates the intra-atomic splitting between
the $\uparrow$- and $\downarrow$-spin states~\cite{Kino},
the effect is very strong and the electronic
structure changes dramatically upon switching between FM, A-, and C-type AFM phases.
Furthermore, the anisotropic AFM ordering has different effect on different $e_g$
orbitals. For example, in the A-type AFM structure, the main lobes of the $x^2$-$y^2$
and $3z^2$-$r^2$ orbitals are aligned correspondingly along FM and AFM bonds.
Since the hopping interactions are penalized
in the case of the AFM coupling,
the A-type AFM arrangement will shrink mainly the $3z^2$-$r^2$ band.
The effect
is reversed in the C-type AFM state, which is associated with the
narrowing of the $x^2$-$y^2$ band.
Thus, the LSDA calculations clearly demonstrate
how the anisotropy of the A- and C-type AFM structures
interplays with the anisotropy of the spacial distribution of
the $x^2$-$y^2$ and $3z^2$-$r^2$ orbitals.

\begin{figure}[h!]
\centering \noindent
\resizebox{12cm}{!}{\includegraphics{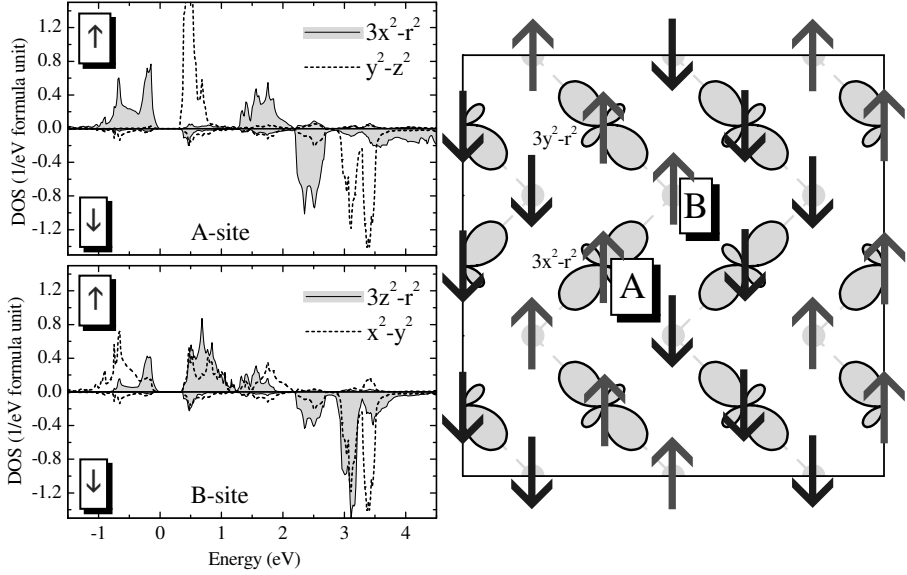}}
\caption{Right panel. Two-dimensional
zigzag antiferromagnetic ordering realized in La$_{1/2}$Sr$_{3/2}$MnO$_4$,
resulting in two different sublattices of the Mn sites
(A and B, typically referred to the ''charge ordering'') and the orbital
ordering (the alternation of partially
populated $3x^2$-$r^2$ and $3y^2$-$r^2$
orbitals at the A sites). Left panel. Corresponding distribution of the
Mn($e_g$) states at the sites A and B in LSDA.
The Fermi level is at zero.}
\label{fig.CEdos}
\end{figure}

  The AFM CE phase presents
the most striking example of such interplay between spin
and orbital degrees of freedom:
the zigzag AFM arrangement appears to be sufficient to explain the
insulating behavior of this phase
(even in LSDA). Therefore, the CE phase can be regarded as a \textit{band insulator},
which is related with the
very peculiar form of the AFM spin ordering (Fig.~\ref{fig.CEdos})~\cite{Lee}.

  The magnetic-state dependence of the electronic structure of the
$e_g$ states is the main microscopic
mechanism, which stands behind the rich variety of the magnetic structures observed
in perovskite manganites. In some sense, these systems show certain tendency to the
self-organization, the main idea of which
can be accumulated in the following formula:
the anisotropic AFM ordering breaks the cubic symmetry of the crystal and leads to the
strong anisotropy of the electronic structure.\footnote{Experimentally, this anisotropy
is frequently observed as an orbital ordering~\cite{Murakami}.}
This anisotropy will be reflected in the anisotropy
of interatomic magnetic interactions (\ref{eqn:Jij}), which depend on the electronic
structure through the matrix elements of the Green function. In many cases,
the response of interatomic magnetic interactions to the change of the
magnetic structure appears to be sufficient to explain the local stability of this
magnetic
state, and
the total energy of such systems will have
many local minima corresponding to different magnetic phases.

\subsubsection{Minimal Model for CMR Manganites}
\label{SubSubSec.DECMRmodel}
  The semi-quantitative model for CMR manganites can be constructed by
using the following arguments.
\begin{itemize}
\item
The model should be based on the realistic electronic structure for the
$e_g$ states and take into account strong dependence of this electronic structure
on the magnetic structure. The proper model for the electronic structure
can be obtained by doing the tight-binding parametrization of
LDA bands~\cite{Priya,OKATB}.
However, for many applications one can use even cruder approximation and consider
only the NN
hoppings between Mn sites, parameterized according to
Slater and Koster~\cite{SlaterKoster}:
$$
t_{\bm{\tau} \bm{\tau} + {\bf n}_\alpha} = \frac{t_0}{2} - \frac{t_0}{2}
\left( \sin \frac{2 \pi \alpha}{3} \sigma_x + \cos \frac{2 \pi \alpha}{3} \sigma_z \right),
$$
where $t_{\bm{\tau} \bm{\tau}'}$$\equiv$$\| t_{\bm{\tau} \bm{\tau}'}^{LL'} \|$ is the
$2$$\times$$2$ matrix in the basis of two $e_g$ orbitals $x^2$-$y^2$ ($L$,$L'$$=$$1$) and
$3z^2$-$r^2$ ($L$,$L'$$=$$2$), and ${\bf n}_\alpha$ is the primitive translation in
the cubic lattice parallel to the $x$ ($\alpha$$=$$1$), $y$ ($\alpha$$=$$2$),
or $z$ ($\alpha$$=$$3$) axis. The anisotropy of the $x^2$-$y^2$ and $3z^2$-$r^2$
orbitals is reflected in the anisotropy of transfer interactions.
The parameter
$t_0$$\propto$$(pd\sigma)^2$ is chosen to reproduce the
$e_g$ bandwidth in LDA ($t_0$$\simeq$$0.7$ eV~\cite{Springer},
and thereafter used as the energy unit).

  The effect of the magnetic structure on the electronic structure is typically included
after transformation to the local coordinate frame, specified by the
directions $\{ {\bf e}_{\bm{\tau}} \}$ of the spin magnetic moments
$$
{\bf e}_{\bm{\tau}} = \left( \cos \phi_{\bm{\tau}} \sin \theta_{\bm{\tau}},
\sin \phi_{\bm{\tau}} \sin \theta_{\bm{\tau}}, \cos \theta_{\bm{\tau}} \right),
$$
and taking the limit of infinite intra-atomic exchange splitting. This yields the
well-known DE Hamiltonian~\cite{MullerHartmann}:
\begin{equation}
{\cal H}^D_{\bm{\tau} \bm{\tau}'} = - \xi_{\bm{\tau} \bm{\tau}'} t_{\bm{\tau} \bm{\tau}'},
\label{eqn:HDE}
\end{equation}
where
$$
\xi_{\bm{\tau} \bm{\tau}'} = \cos \frac{\theta_{\bm{\tau}}}{2} \cos \frac{\theta_{\bm{\tau}'}}{2}
+ \sin \frac{\theta_{\bm{\tau}}}{2} \sin \frac{\theta_{\bm{\tau}'}}{2}
e^{-i (\phi_{\bm{\tau}} - \phi_{\bm{\tau}'})}
$$
describes modulations of transfer interactions caused by deviations from the
FM spin alignment.\footnote{
The DE Hamiltonian is essentially non-local with respect to the site indices.
In this sense one can say that
the DE physics is mainly the \textit{physics of bonds}, which may have a direct
implication to the behavior of paramagnetic phase of CMR manganites~\cite{PRB03}.}
\item
${\cal H}^D_{\bm{\tau} \bm{\tau}'}$ is proportional to $t_0$, and similar to the
DE term considered in Sec.~\ref{SubSec.modelDE}, describes the FM interactions
in the systems, which are now generalized to the case of arbitrary spin
arrangement.
This term is combined with the energy of phenomenological
AFM SE interactions:
$$
{\cal H}^S_{\bm{\tau} \bm{\tau}'} = -\frac{1}{2} J^S {\bf e}_{\bm{\tau}} \cdot {\bf e}_{\bm{\tau}'},
$$
where $J^S$$<$$0$ is restricted by the
nearest neighbors and contains contributions of both $t_{2g}$ and $e_g$ states.
Using an analogy with the SE model considered in Sec.~\ref{SubSec.modelSE}, it is assumed that
the exchange constant $J^S$ does not depend on the magnetic state, and all such dependencies come
exclusively
from the DE term.
\end{itemize}

\begin{figure}[h!]
\centering \noindent
\resizebox{12cm}{!}{\includegraphics{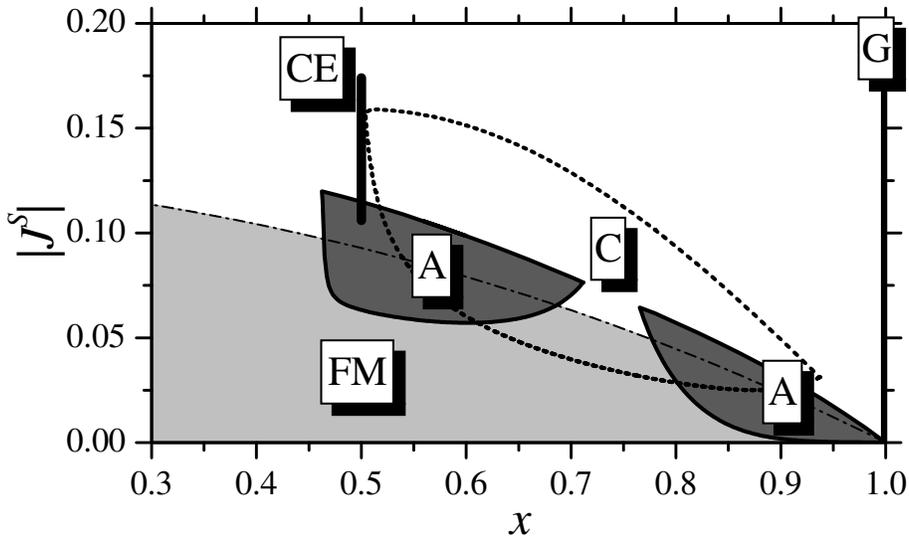}}
\caption{The areas of local stability of the main collinear phases with respect
to the spin canting in the degenerate double exchange model (from Ref.~\protect\cite{PRB01}).}
\label{fig.DEphase}
\end{figure}

  The combination of ${\cal H}^D_{\bm{\tau} \bm{\tau}'}$ and ${\cal H}^S_{\bm{\tau} \bm{\tau}'}$
constitutes the minimal model, which explains the origin of the main magnetic structures
observed in doped manganites.\footnote{The on-site Coulomb interaction~\cite{Maezono}
and the lattice distortion~\cite{Fang}
have been also considered as the model ingredients. However, as far as the low-temperature
behavior is concerned, the inclusion of these terms (of course, under the appropriate choice of
the parameters~\cite{Shen}) may change the situation quantitatively, but not qualitatively.}
An example of theoretical phase diagram is shown in
Fig.~\ref{fig.DEphase}~\cite{BrinkKhomskii,PRB01,Yunoki}.
The position and the order of the main magnetic phases with respect to the hole-doping
has clear similarity with the experimental phase diagram (Fig.~\ref{fig.NSMO}).\footnote{
Note that around $x$$=$$0.5$, $|J^S|$ is of the order of 0.1~\cite{Springer}.}

  The origin of the CE-type AFM phase requires special comment, because
canonically it was attributed to the charge ordering of the Mn$^{3+}$ and
Mn$^{4+}$ ions~\cite{Goodenough}, and this point of view was dominating
for many years. The modern view on the problem is that CE is the regular
magnetic phase, whose properties can be understood in the degenerate DE model
similar to other magnetic states~\cite{Springer,Dagotto,CEpapers}.
The insulating behavior of this phase is related with the topology
of one-dimensional zigzag FM chains.\footnote{
The $e_g$ electron passing through the $3z^2$-$r^2$ and $x^2$-$y^2$ orbitals
of the corner (B) sites (Fig.~\ref{fig.CEdos})
correspondingly preserves and changes its phase.
In the one-dimensional case, it will open the band gap
correspondingly at the boundary
and in the center of the Brillouin zone, that explains the
insulating behavior.}
This also explains why the CE phase exists in the very narrow region
close to $x$$=$$0.5$, i.e. when the Fermi energy falls into the band gap.
More generally, any periodic one-dimensional zigzag object composed of the
$e_g$ orbitals will behave as a band insulator at certain commensurate
values of the electron density~\cite{Springer,Hotta}.

\subsubsection{Implication to the CMR Effect}
\label{SubSubSec.DECMRCMR}
  The magnetic origin of the CE phase has many implications to the CMR effect
and readily explains why the insulating state of some manganites can be easily
destroyed by the external magnetic field. However, it would not be right to say
that CMR is exclusively related with the existence of the CE phase.
It is a more general phenomenon inherent to the DE physics. Below we will
consider two such possibility.

  1. According to
the form of the DE Hamiltonian (\ref{eqn:HDE}),
the transfer interactions are forbidden between
antiferromagnetically coupled sites. Therefore, from the viewpoint of
transport properties,
the A-type AFM phase
behaves as an insulator in the $z$-directions and as a metal
in the $xy$-plane. The A state can be continuously transformed to the
spin-canted state~\cite{PRB01} by applying the external magnetic field.
The spin-canting unblocks the hopping interactions and
gradually decreases the
resistivity along $z$. This is the basic idea of
the so-called ''spin-valve'' effect,
observed in the A-type AFM
Nd$_{0.45}$Sr$_{0.55}$MnO$_3$~\cite{SpinValve}.

  The unique aspect of the CE phase is that it has a band gap, and therefore is insulating
in all the directions, including the direction of propagation of the FM zigzag chains. This
predetermines rather distinguished behavior of the CE phase in the magnetic field.
As it was pointed out before, the stability of the CE phase is directly related with
the existence of the band gap. Similar to the A-type AFM phase, the external magnetic field
leads to the canting of spins also in the CE phase. The basic
difference is that starting with certain angle
such canting will close the band gap.\footnote{In the two-dimensional DE model,
the band gap is closed when the angle between
spin magnetic moments in adjacent zigzag chains becomes smaller than 110$^\circ$.}
Then, the CE phase will become unstable and the system will abruptly
transform to the metallic FM phase, which is favored by the magnetic field.
The phenomenon is well know as the ''melting of the charge-ordered state'', which
accompanied by the abrupt drop of the resistivity~\cite{GordonBreach}.

\begin{figure}[h!]
\centering \noindent
\resizebox{5cm}{!}{\includegraphics{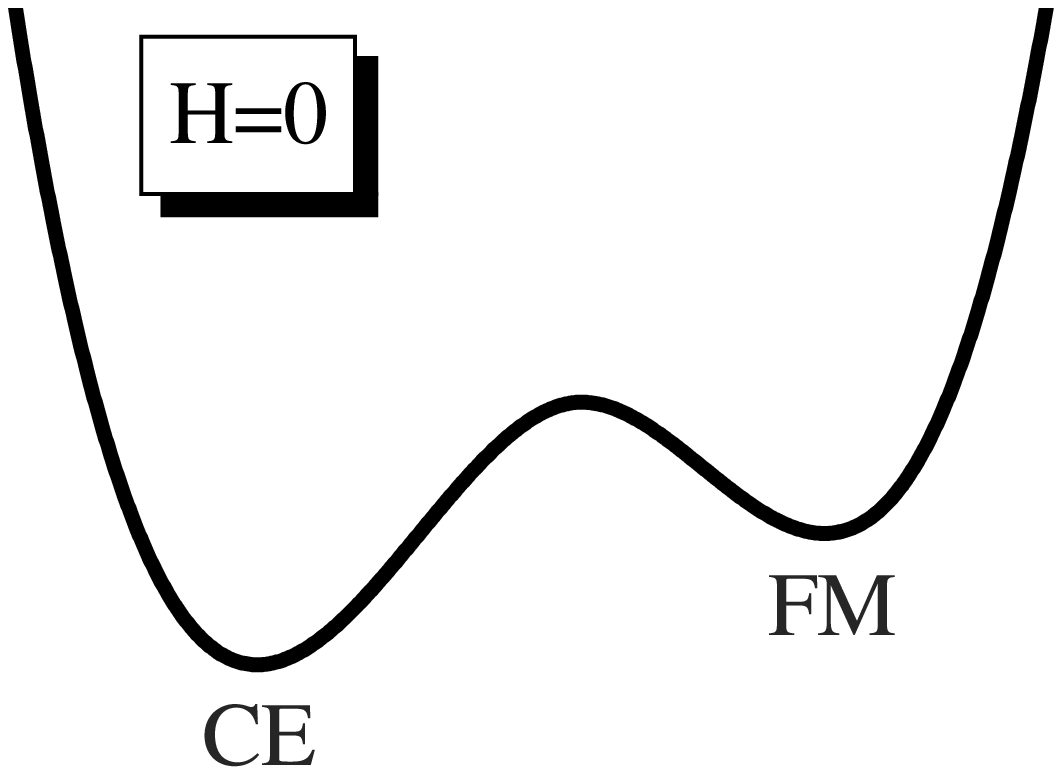}}
\resizebox{5cm}{!}{\includegraphics{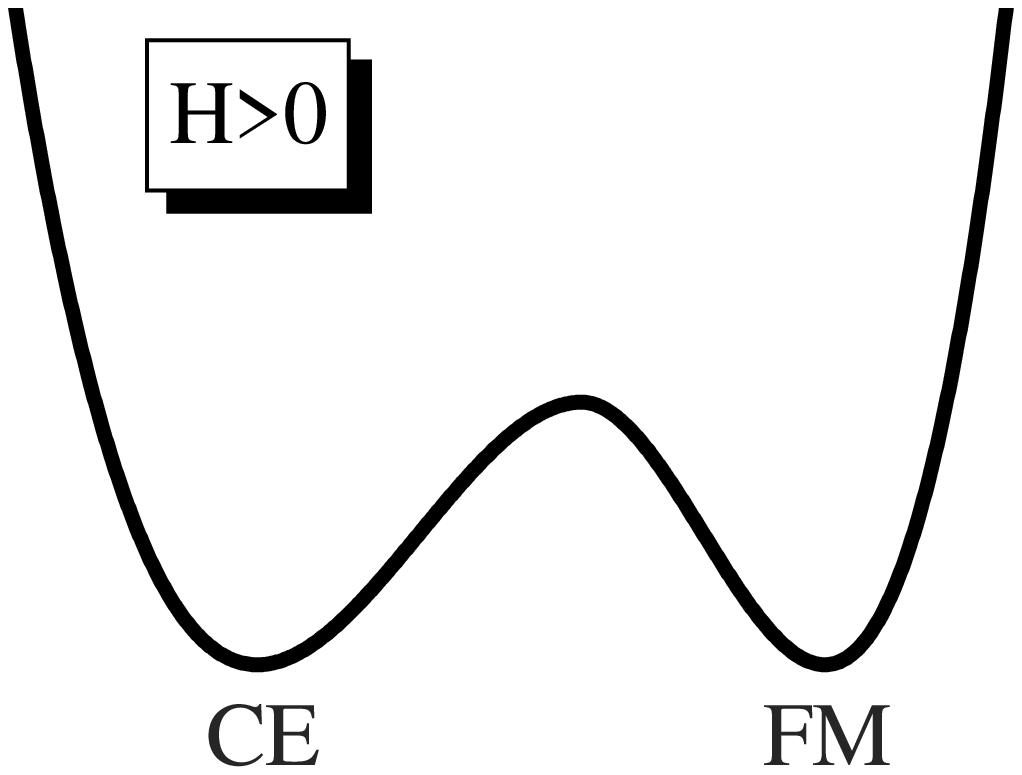}}
\\
\fbox{\resizebox{5cm}{!}{\includegraphics{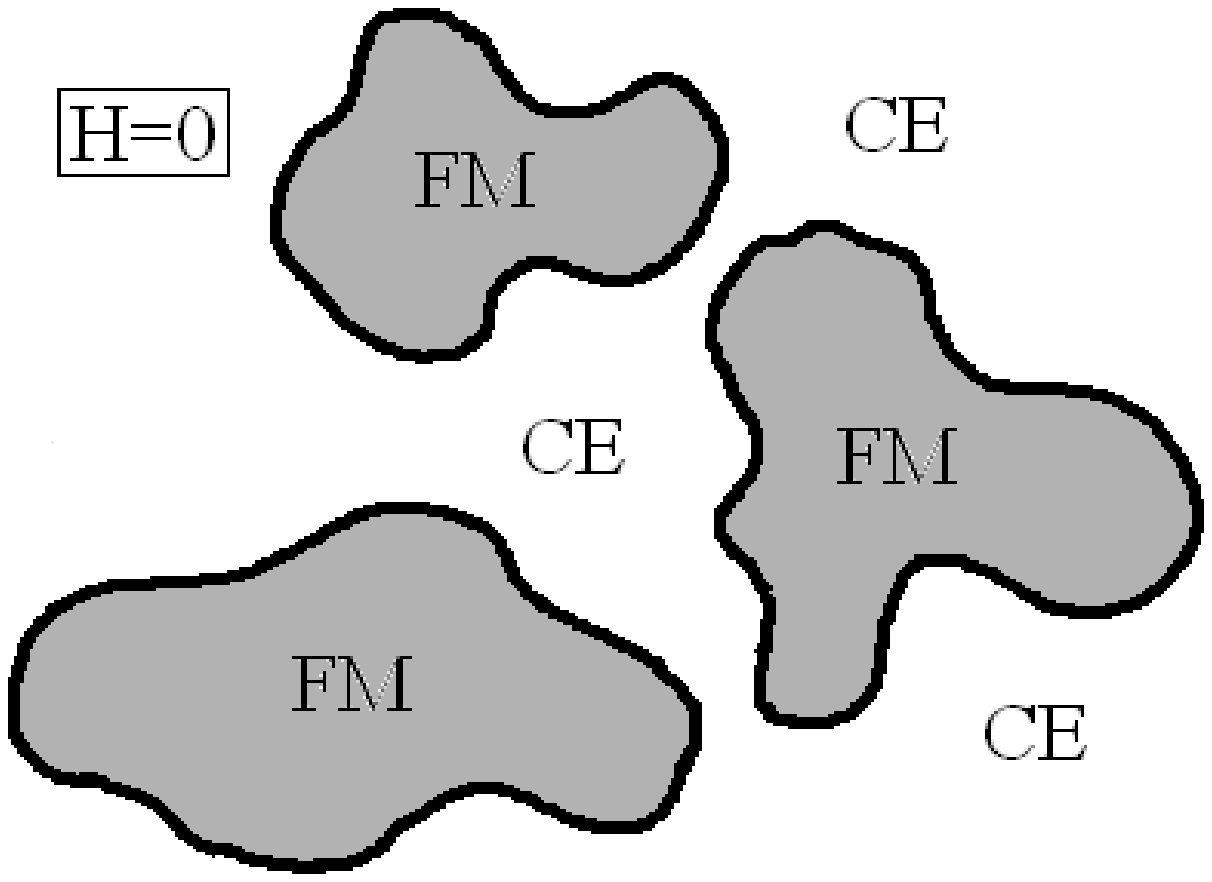}}}
\fbox{\resizebox{5cm}{!}{\includegraphics{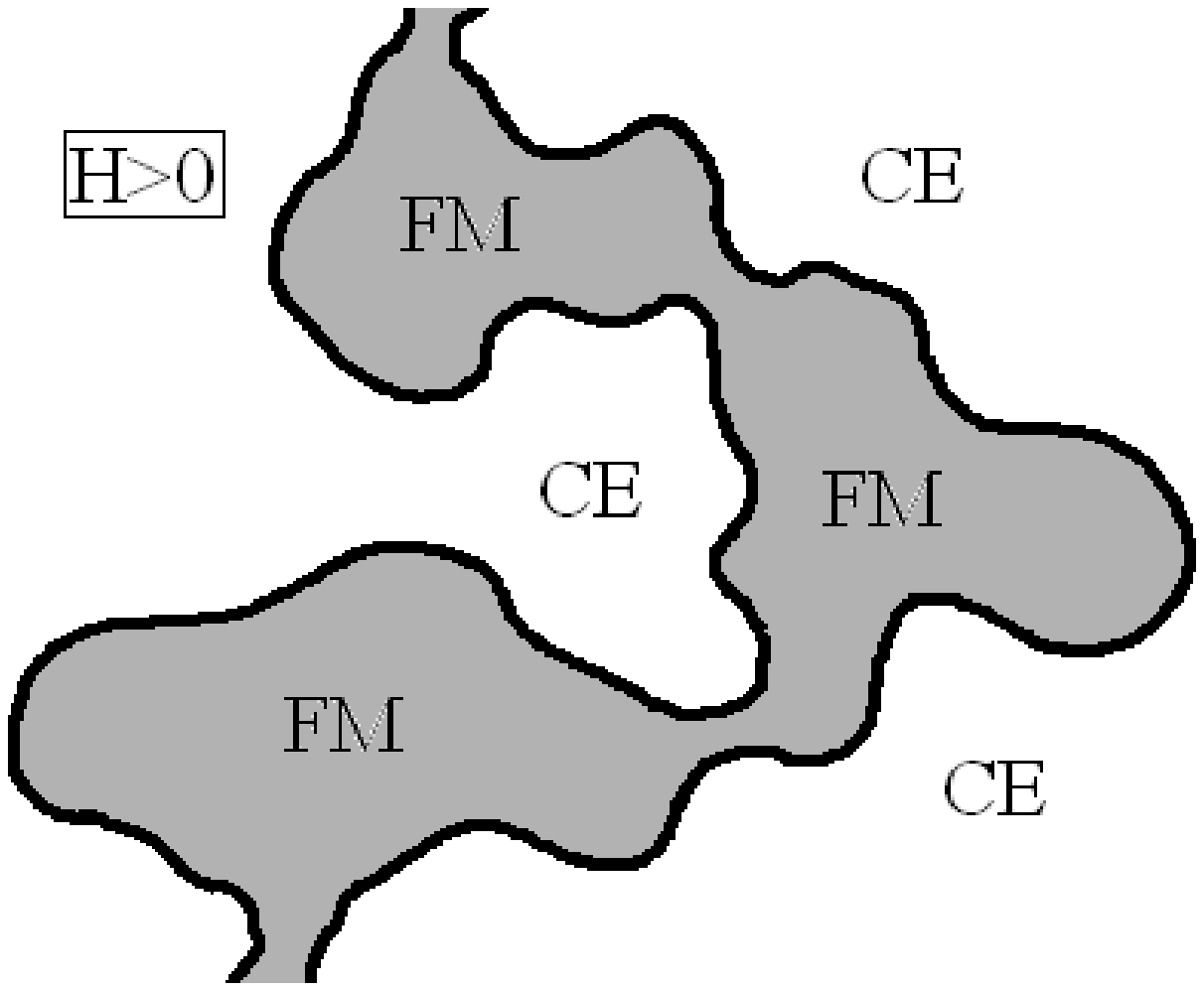}}}
\caption{Schematic view on the percolative scenario of the CMR effect.
Top: Local minima of the double exchange energy corresponding to the
insulating CE and metallic FM phases. The relative position of these
minima is controlled by the magnetic field $H$.
Bottom: mixed-phase state realized at the finite temperature. The magnetic field
increases the amount of the FM phase, and
above the percolation threshold forms the conducting FM path.}
\label{fig.percolation}
\end{figure}

  2. Typical theories of the second type are based on the following observations~\cite{Moreo}:\footnote{
The percolative scenario of the CMR effect in manganites has been originally proposed by
Nagaev~\cite{Nagaev}. The newer theories provide a quantitative description of this effect
and clarify the origin of the main magnetic states,
particularly -- the existence of the insulating CE-type AFM phase in the degenerate DE model.}
\begin{itemize}
\item
The total energy of the DE model may have several local minima corresponding
to different FM and AFM phases, which may exist in the same range of the
hole-doping. This also implies the first-order character of the phase boundaries~\cite{Yunoki}.
The concrete example is the region close to $x$$=$$0.5$ shown
in Fig.~\ref{fig.DEphase}.
\item
Some of these phases are insulating due to the special (zigzag) geometry of the
AFM pattern.
\end{itemize}
The situation is schematically illustrated in Fig.~\ref{fig.percolation}.
At $T$$=$$0$ and without field, the system is trapped in the insulating CE-type AFM
ground state. The next minimum, corresponding to the FM metallic state should be located
within the energy range accessible by the external magnetic field.
At the finite temperature and neglecting interactions between different phases, the system will
be represented by the mixture of these two states, due to the configuration mixture
entropy.\footnote{The appearance of the two-phase state can be considerable
facilitated in the presence of impurities and the cation disorder~\cite{Moreo,Nagaev}.
As we will argue in Sec.~\ref{SubSec.DoubleP}, this effect plays a very
important role in the double perovskites.}
The relative amount of these states will depend on the temperature $T$ and the
magnetic field $H$. The latter controls the relative position of the CE and FM minima.
Without field, the CE phase will dominate, while the FM phase will form metallic
islands in the insulating sea. The application of the magnetic field will increase
the amount of the of the FM phase. At some point, which is called the percolation
threshold, the FM islands become connected by forming the conducting FM paths
throughout the spacemen. This will correspond to the sharp drop of the resistivity.

  Since many interesting (and, presumably, the most practical) phenomena of
perovskite manganites are related with the existence of zigzag AFM structures,
it is very important to understand the behavior of these structures, and especially
the way how they will evolve with the temperature, in the magnetic field, or
upon the change of the crystal structure. We would like to close this section
by listing some interesting and not completely resolved problems in this direction.
\begin{itemize}
\item
The existence of two transition temperatures, $T_{\rm N}$ and $T_{\rm CO}$,
which are typically attributed to the onset of N\'{e}el
AFM and charge ordering, respectively.
In reality, both transitions are probably magnetic and $T_{\rm N}$ corresponds to the
order-disorder transition which takes place only between FM zigzag chains and
largely preserves the coherent motion of spins within the chains, while
$T_{\rm CO}$ corresponds to the transition to the totally
disordered PM state~\cite{condmat2002}.
\item
Appearance of incommensurate charge and orbital superstructures just below
$T_{\rm CO}$~\cite{COincommensurate}.
\item
Effects of cation disorder and appearance of new zigzag superlattices upon
artificial ordering of $R$ and $D$ elements in some three-dimensional
manganites~\cite{Arima}.
\end{itemize}

\subsubsection{Role of Rare-earth, Alkaline-earth, and Oxygen and States}
\label{SubSubSec.OLa}
  In this section we will briefly discuss the influence of the $R$, $D$, and
oxygen states on the interatomic magnetic interactions in $R_{1-x}D_x$MnO$_3$
in the context of revisions in the form of these interactions, which were
recently considered by Bruno~\cite{Bruno}. We will concentrate on the behavior of
the cubic FM phase of the virtual-crystal alloy La$_{0.7}$Ba$_{0.3}$MnO$_3$,
and treat the problem of interatomic magnetic interactions using the frozen
spin wave approximation supplemented with the LSDA.\footnote{
The cubic lattice parameter is $a_0$$=$$3.876$\AA.}
In the FM phase, the magnetic Mn sites polarize neighboring La/Ba and oxygen
sites. The typical distribution of the magnetic moments amongst different
sites is $\mu_{\rm Mn}$$=$$3.48$, $\mu_{\rm La/Ba}$$=$$0.08$, and
$\mu_{\rm O}$$=$$0.05$ $\mu_B$. Our main concern will be the role played by
$\mu_{\rm La/Ba}$ and $\mu_{\rm O}$ in the spin dynamics of La$_{1-x}$Ba$_x$MnO$_3$.

\begin{figure}[h!]
\centering \noindent
\resizebox{10cm}{!}{\includegraphics{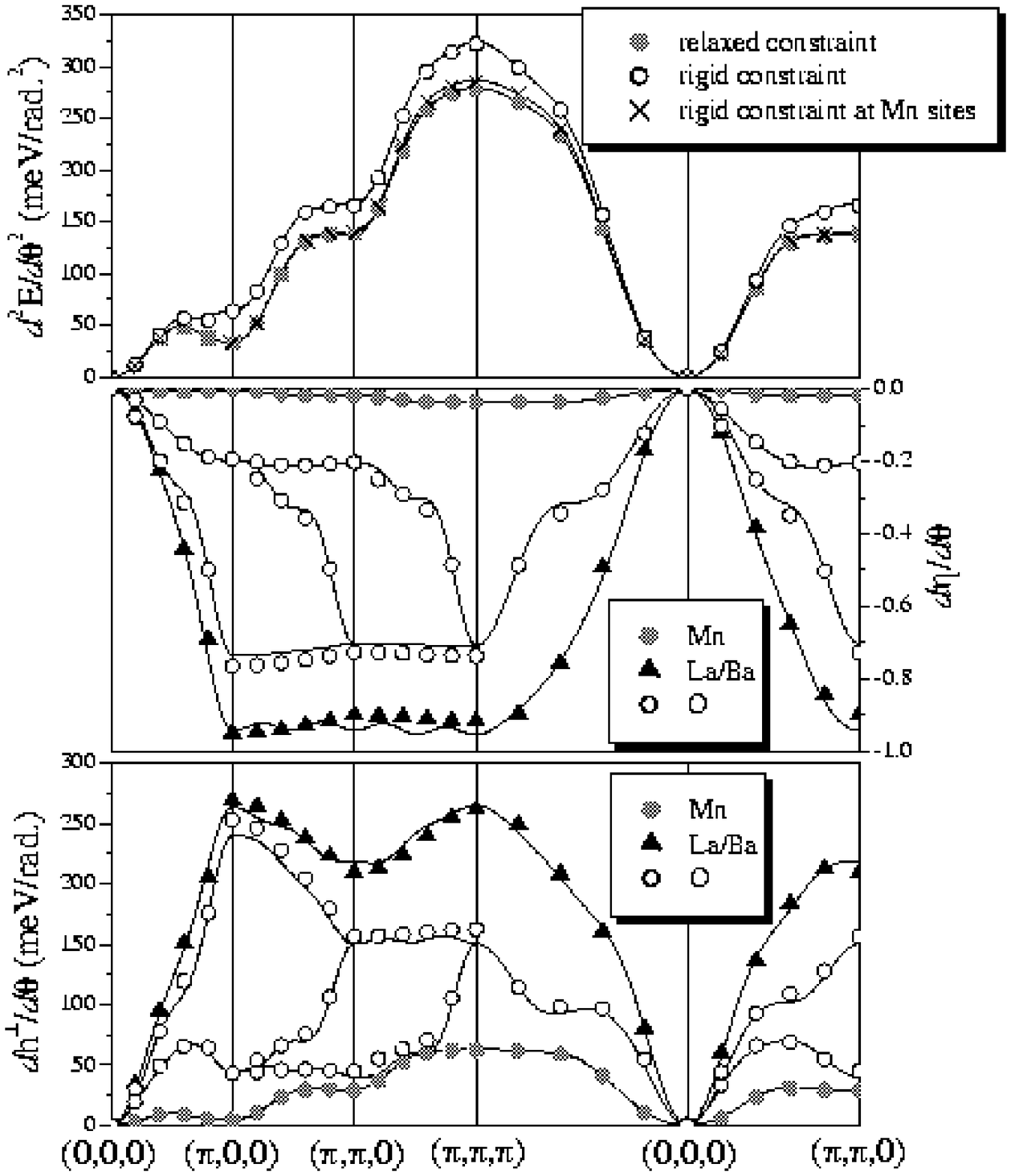}}
\caption{Results of frozen spin wave calculations for La$_{0.7}$Ba$_{0.3}$MnO$_3$:
second derivatives of the total energy (which specify the form of
the magnon dispersion) obtained by
rotating the Kohn-Sham effective fields (relaxed constraint) and by applying
additional perpendicular constraining fields $\{ {\bf h}^\perp_{\bm{\tau}} \}$
(rigid constraint) at all sites and the Mn sites only; the parasitic canting of the
magnetic moments in the relaxed constraint calculations; and the perpendicular
magnetic field applied in order to compensate this canting. Symbols show the
calculated points, and the solid lines are results of an interpolation.}
\label{fig.SWLBMO}
\end{figure}

  We assume that the cone-angle of the spin wave does not
depend on the atomic site: $\theta_{\bm{\tau}}$$\equiv$$\theta$. The phase
of the spin wave is modulated as ${\bf q} \cdot \bm{\tau}$ with
$\bm{\tau}$ running over all Mn, O, and La/Ba sites. Then, for each ${\bf q}$
we calculate the second derivative of the total energy with respect to $\theta$
using the magnetic force theorem. The derivative
can be also used to estimate the magnon spectrum in the FM state:
$$
\omega_{\bf q}=\frac{2}{\mu} \left. \frac{d^2 E({\bf q},\theta)}{d \theta^2} \right|_{\theta = 0}
$$
($\mu$ being the FM moment). We consider two possibility:
\begin{figure}[h!]
\centering \noindent
\resizebox{4cm}{!}{\includegraphics{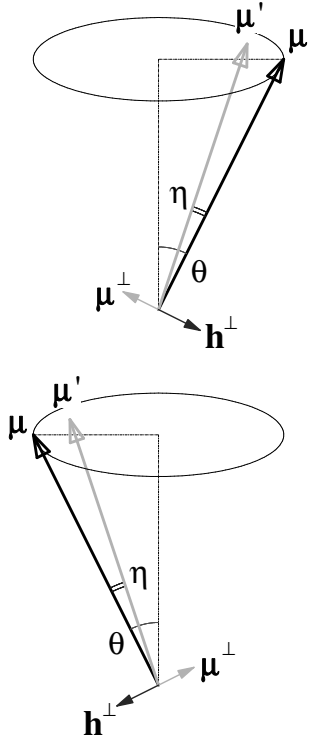}}
\caption{Typical situation realized in frozen spin wave calculations for ${\bf q}$$=$$(0,0,\pi)$.
$\{ \bm{\mu} \}$ is the constrained distribution of the magnetic moments.
$\{ \bm{\mu}' \}$ is the distribution of magnetic moments obtained after solution
of Kohn-Sham equations with the rotated effective fields.
Generally, the spins $\{ \bm{\mu}' \}$ are canted towards the ground state configuration
by the angle $\eta$, and have the components $\{ \bm{\mu}^\perp \}$, which are perpendicular to
$\{ \bm{\mu} \}$.
The constraining fields $\{ {\bf h}^\perp \}$
are included in order to compensate $\{ \bm{\mu}^\perp \}$.}
\label{fig.SW}
\end{figure}
\begin{itemize}
\item
Rigid constraint, obtained by rotating the KS effective field and enforcing the
directions of the spin magnetic moments at each site to follow exactly the
form of the spin wave. The latter is achieved by applying
the perpendicular
constraining fields $\{ {\bf h}^\perp_{\bm{\tau}} \}$ (Fig.~\ref{fig.SW}), which can be estimated using the
response-type arguments~\cite{Bruno}.
The fields $\{ {\bf h}^\perp_{\bm{\tau}} \}$ affect the KS orbitals $\{ \psi_i \}$,
and through the change of these orbitals,
correct the form of the spin-magnetization density (\ref{eqn:sdensity}).
However, they do not contribute explicitly
to the total energy change (\ref{eqn:LFT}).
\item
Relaxed constraint, obtained after rotation of the KS effective fields.
\end{itemize}
Results of these calculations are shown in Fig.~\ref{fig.SWLBMO}.
The magnon dispersion has a characteristic form, which is
manifested in the pronounced softening
at the Brillouin zone boundaries~\cite{softening}.\footnote{
Experimentally, the softening of the spin-wave dispersion is
frequently accompanied by a broadening, though these two effects are
not necessarily related with each other and the materials which
show very similar softening of the magnon spectrum can reveal very
different broadening~\cite{softening}.}
The origin of this softening is related with the behavior of interatomic
magnetic interactions, and can be understood on the basis of
realistic electronic structure calculations~\cite{PRL99}.
It has a direct implication to the stability of the FM state and the form
of the magnetic phase diagram versus the hole-doping~\cite{Springer}.

  In this section, our main concern will be a little bit different. As it is seen
in Fig.~\ref{fig.SWLBMO}, the application of the constraining fields
$\{ {\bf h}^\perp_{\bm{\tau}} \}$ can significantly affect the magnon dispersion.
For example, around $(\pi,0,0)$, $\omega_{\bf q}$'s calculated using two different
schemes can differ by factor two.
The most interesting question is where this
difference comes from. The simple rotation of the KS effective fields by the angle
$\theta$ causes the parasitic canting of spins off the
constrained directions by the angles
$\{ \eta_{\bm{\tau}} \}$, which are proportional to
$\theta$ if the latter is small (see Fig.~\ref{fig.SW}). However, this canting can be very different
at different magnetic sites. Indeed, $\eta_{\bm{\tau}}$ is of the order of
$\theta$ at the La/Ba and oxygen sites, and almost negligible at the Mn sites.
Therefore, the fields $\{ {\bf h}^\perp_{\bm{\tau}} \}$, which are introduced in order to correct
this canting will be the largest at the La/Ba and oxygen sites, and the smallest at the Mn sites.
An attempt to apply ${\bf h}^\perp_{\bm{\tau}}$ only at the Mn sites has a negligible effect
on the magnon dispersion. These results rise several important questions, which still need to
be understood.

  $\bullet$ There is a large
intrinsic uncertainty in the behavior of spin magnetization at the La/Ba and oxygen sites,
and especially in the form, in which these sites contribute to the spin dynamics of
La$_{1-x}$Ba$_x$MnO$_3$ and similar compounds. The problem may be directly related with the
definition of the adiabaticity concept, underlying the frozen spin wave approach. According
to this concept, the magnetic systems have two characteristic time
scales. The spin directions are typically regarded as the slow variables, while the
internal electronic degrees of freedom are the fast ones, which self-consistently follow
the instantaneous orientational distribution of the spin-magnetization density.
From this point of view, the La/Ba and oxygen sites have clearly a dual character.
On the one hand, neither of them can be regarded as the source of magnetism in these systems.
On the other hand, both of them carry a significant weight of the spin-magnetization density,
mainly due to the hybridization with the magnetic Mn sites. Then, how the
La/Ba and oxygen states shall be treated? In principle, one may have several possibilities.
\begin{enumerate}
\item
The La/Ba and oxygen sites are magnetic and shell be treated on the same level as the
Mn sites, by applying independent constraint conditions for the slow-varying directions
of the La/Ba and oxygen spins. This would correspond to the appearance of additional
magnon branches in the frozen spin wave calculations. Clearly, this scenario can be
sorted out from the beginning, because both La/Ba and oxygen moments
are induced by the hybridization with the Mn sites and cannot be totally independent from
the latter.
\item
The contribution of La/Ba and oxygen states is purely electronic, which implies that these
degrees of freedom are ''fast'' and self-consistently follow the distribution of the
spin magnetic moments at the Mn sites. This would roughly correspond to the relaxed
constraint conditions in the spin-wave calculations shown in Fig.~\ref{fig.SWLBMO}.
\item
An intermediate situation when the electronic degrees of freedom at the
La/Ba and oxygen sites are not ''sufficiently fast'' so that the magnon dispersion
shall be averaged over several intermediate configurations of the La/Ba and oxygen spins.
This is the most interesting scenario, which is however
only a speculation. If true, the magnon dispersion in FM manganites should have
an intrinsic bandwidth, which should be of the same order of magnitude as the
difference between ''rigid'' and ''relaxed'' calculations shown in Fig.~\ref{fig.SWLBMO}.
In fact, the strong magnon broadening has been observed experimentally in several
FM manganites~\cite{softening}, though it is of course premature to make a
direct connection between this broadening and the behavior of La/Ba and oxygen states
considered in this section.
\end{enumerate}

  $\bullet$ The non-collinear distribution of magnetic moments at the
La/Ba and oxygen sites can interfere with other (extrinsic) factors,
such as randomness effects caused by the cation disorder. Then, the chemical
disorder may lead to the orientational spin disorder at the
La/Ba and oxygen sites, which may also contribute to the magnon broadening
associated with the randomness~\cite{Motome}.

\subsection{Double Perovskites}
\label{SubSec.DoubleP}

  Ordered double perovskites, like Sr$_2$FeMoO$_6$ and Sr$_2$FeReO$_6$,
is another interesting class of compounds~\cite{Kobayashi}.
They exhibit fairly large magneto-resistance effect, which coexists with
relatively high magnetic transition temperature (415 and 401 K for
Sr$_2$FeMoO$_6$ and Sr$_2$FeReO$_6$, respectively). The latter factor presents a great
advantage over CMR manganites, and opens a way to technological applications
of these compounds in devices operating at the room temperature.

\begin{figure}[h!]
\centering \noindent
\resizebox{8cm}{!}{\includegraphics{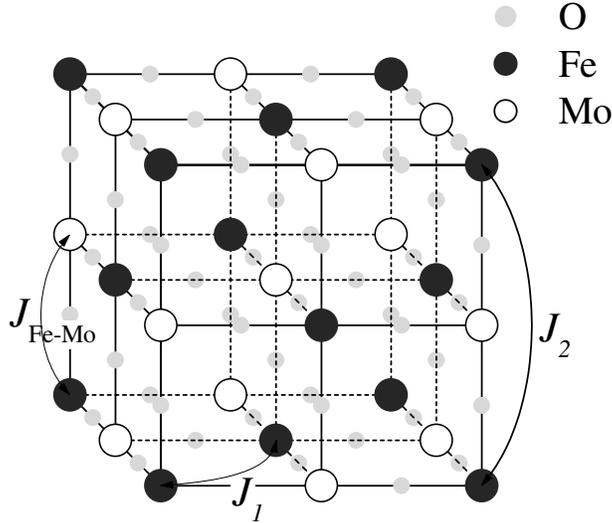}}
\caption{Positions of Fe, Mo, and oxygen sites in ordered
double perovskite Sr$_2$FeMoO$_6$. Arrows show the main magnetic
interactions responsible for properties of this compound.}
\label{fig.SFMOlattice}
\end{figure}

  At the first sight, the double perovskites crystallize in a simple
crystal structure. It can be regarded as the cubic SrFeO$_3$, in which every
second Fe is replaced by Mo or Re (Fig.~\ref{fig.SFMOlattice}).
However, there are two fundamental problems:
\begin{enumerate}
\item
\textit{Anti-site defects.} Typically, certain percent of atoms from the
nominally Fe sublattice interchanges with the same amount of atoms
from the Mo sublattice. The number of such defects in the best available
single-crystalline samples of Sr$_2$FeMoO$_6$ is about 8\%~\cite{Tomioka}.
\item
\textit{Fine details of the crystal structure.} Typically, the information about
directions and
magnitudes of the oxygen displacement is either unknown or rather
controversial.
\end{enumerate}

  The extraordinary properties of Sr$_2$FeMoO$_6$ and Sr$_2$FeReO$_6$ are
usually attributed to the half-metallic electronic structure expected
in the FM state. In this sense,
the double perovskites are typically considered as an example of relatively simple
and well established physics.

  In this section, we will discuss the behavior of interatomic magnetic interactions
in Sr$_2$FeMoO$_6$ and argue that this point of view must be largely revised.
Particularly, we will show that the half-metallic electronic structure is
incompatible with stability of the FM ground state.
Therefore, there should be some additional mechanism which stabilizes the
FM state and presumably destroys the half-metallic character of the
electronic structure. We propose that the unique properties of the double
perovskites are directly
related with the above-mentioned
defects and uncertainties of the crystal structure~\cite{PRB02}.\footnote{The same
arguments can be applied to Sr$_2$FeReO$_6$.}

\begin{figure}[h!]
\centering \noindent
\resizebox{12cm}{!}{\includegraphics{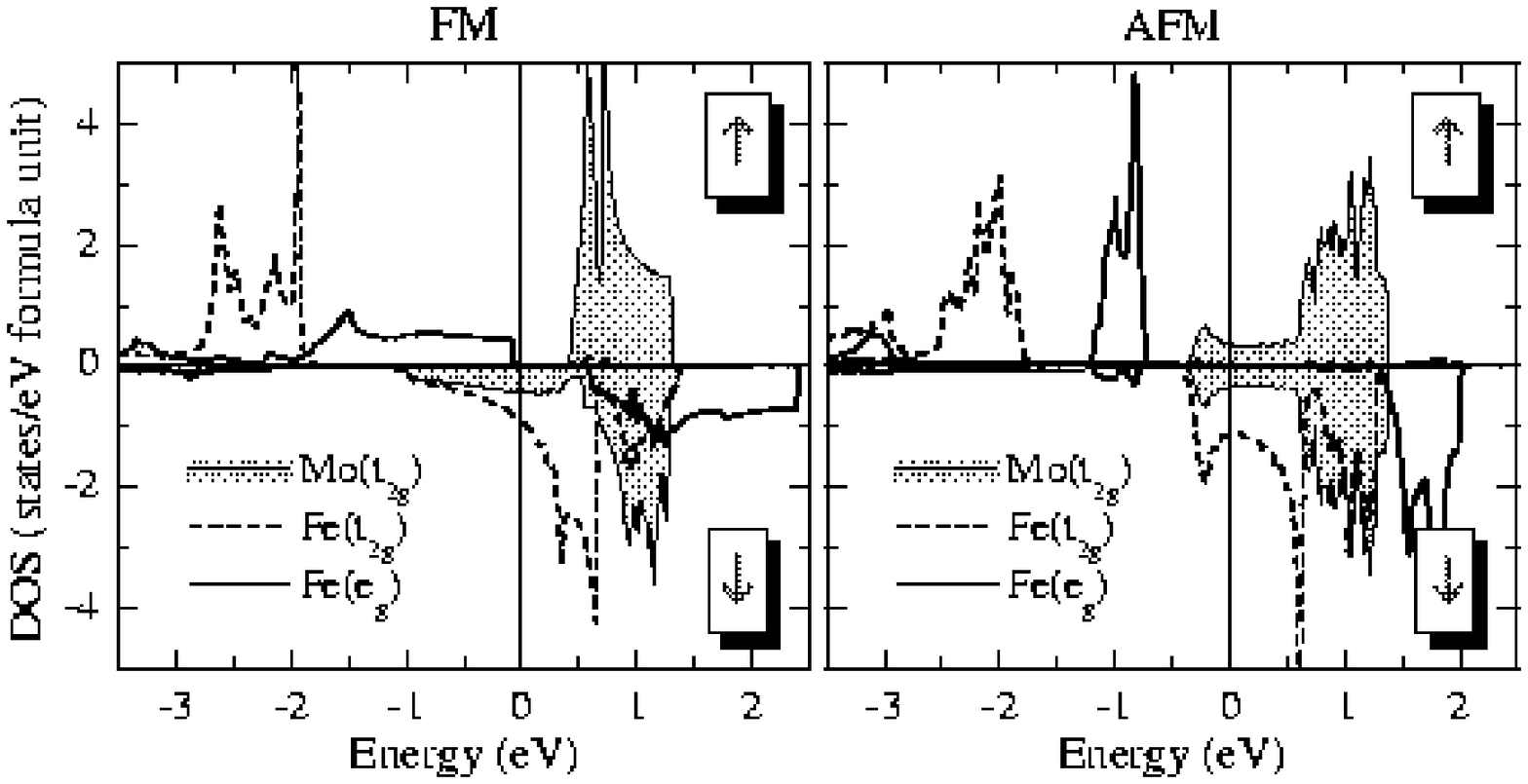}}
\caption{Local densities of states corresponding to
the ferromagnetic (left) antiferromagnetic (right) alignment of Fe spins in
ordered
Sr$_2$FeMoO$_6$, in LSDA. The Fermi level is at zero.}
\label{fig.SFMOdos}
\end{figure}

  A rough idea about the problem can be obtained from rather simple considerations
of the electronic structure of Sr$_2$FeMoO$_6$ (Fig.~\ref{fig.SFMOdos}),
which suggests that
similar to the CMR manganites
there are two competing interactions which define the
form of the magnetic ground state in this compound.
\begin{enumerate}
\item
It is true that the electronic structure of the hypothetical FM phase of
Sr$_2$FeMoO$_6$ is half-metallic: the Fermi level crosses the
hybrid $t_{2g}$ band in the $\downarrow$-spin channel and falls in
the gap between Fe($e_g$) and Mo($t_{2g}$) bands in the
$\uparrow$-spin channel. The partly-filled $\downarrow$-spin
$t_{2g}$ band will generally be the source of FM interactions,
due to the double exchange mechanism considered in
Sec.~\ref{SubSec.modelDE}, which will be somewhat
modified by the fact that the transfer
interactions between spin-polarized Fe($t_{2g}$) orbitals are indirect
and operate via nonmagnetic Mo($t_{2g}$) states~\cite{KanamoriTerakura}.
This interaction will also polarizes
the Mo($t_{2g}$) states antiferromagnetically with respect to the
Fe sublattice.
\item
The half-filled Fe($e_g$) states should generally contribute to
AFM
interactions. Typically, this contribution is expected to be small,
because in the double-perovskite structure, the neighboring Fe sites
are separated by the long O-Mo-O paths. A very rough idea about the
strength of these interactions can be obtained from the comparison of
densities of the Fe($e_g$) states in the
FM and (type-II~\cite{Terakura}) AFM phases:
if these interaction were weak, the bandwidth
would not depend on the magnetic state.
However, the direct calculations show the opposite trend and the
Fe($e_g$) band considerably shrinks in the case of the AFM alignment.
\end{enumerate}

  A simple estimate of the relative strength of the FM and AFM
contributions, irrespective on their decomposition onto the
interatomic magnetic interactions, can be obtained from the
one-electron (band) energies
$$
E_b = \int \varepsilon
{\cal N}(\varepsilon) d \varepsilon
$$
(${\cal N}$ being the total density of states) calculated separately
for the $e_g$ and $t_{2g}$ bands in the FM and AFM phases.
These energies are listed in Table~\ref{tab:EbSFMO}.\footnote{
Note that in the case of SE interactions, in addition to the change of
the Fe($e_g$) band energy, there will be an additional contribution
associated with the shift of the O($2p$) states
(Sec.~\ref{SubSec.modelSE} and Ref.~\cite{Oguchi}).}
\begin{table}[h!]
\caption{One-electron energies associated with the hybrid $t_{2g}$ band
located near the Fermi level and the occupied Fe($e_g$) band in the FM and AFM
states of Sr$_2$FeMoO$_6$.}
\label{tab:EbSFMO}
\begin{center}
\begin{tabular}{ccc}\hline
band      & FM              & AFM           \\\hline
$t_{2g}$  & -1.8            & -1.7          \\
$e_g$     & -4.6            & -4.7          \\\hline
\end{tabular}
\end{center}
\end{table}
One can clearly see that in the AFM state, the energy loss associated with the
hybrid $t_{2g}$ band is totally compensated by the energy gain associated with
the Fe($e_g$) band. This example shows that the problem is indeed very subtle
and the half-metallic electronic structure obtained in the FM state does not
necessary guarantee that this state will be locally stable and realized as the
ground state. It does not explain the high value of the
magnetic transition temperature either.

  The conclusion is supported by direct calculations of interatomic magnetic
interactions using both frozen spin wave and Green's function methods.

\begin{figure}[h!]
\centering \noindent
\resizebox{12cm}{!}{\includegraphics{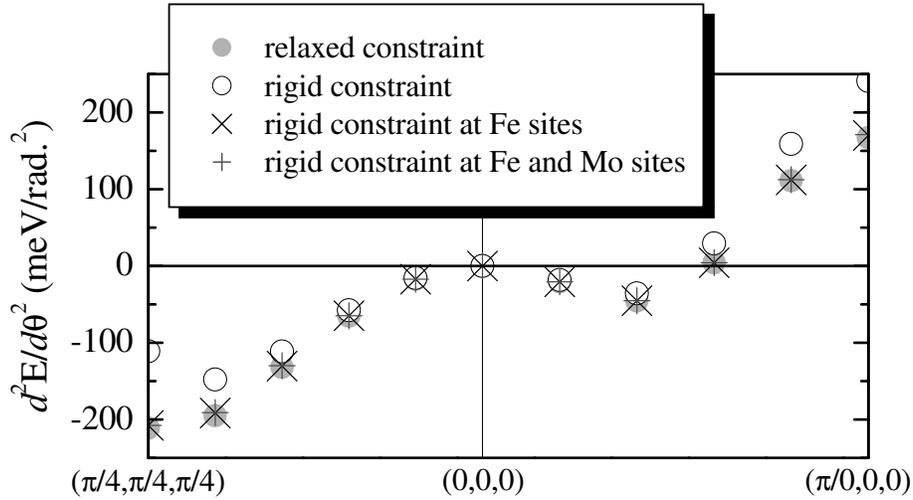}}
\caption{Results of frozen spin wave calculations for ordered Sr$_2$FeMoO$_6$
(see Fig.~\protect\ref{fig.SWLBMO} for details).}
\label{fig.SFMOsw}
\end{figure}

  1. The FM state appears to be unstable with respect to the non-collinear
spin-spiral alignment realized at certain ${\bf q}$'s, for which
$\frac{d^2}{d \theta^2} \left. E({\bf q},\theta) \right|_{\theta=0}$$<$$0$
(Fig.~\ref{fig.SFMOsw}).
The second derivative of the total energy
was calculated using several different approximations, similar to
the analysis of the spin-wave dispersion of
La$_{0.7}$Sr$_{0.3}$MnO$_3$ undertaken in Sec.~\ref{SubSubSec.OLa}:
by imposing the rigid constraint on the directions of spin
magnetic moments at all sites of Sr$_2$FeMoO$_6$, and only at the Fe
and Mo ones. Similar to the La$_{0.7}$Sr$_{0.3}$MnO$_3$ case,
the main effect of the constraining fields $\{ {\bf h}^\perp_{\bm{\tau}} \}$
is associated with the Sr and oxygen sites, whereas the contributions
of the Fe and Mo sites are negligible.
The ''rigid constraint'' can be regarded as the most consistent check for the
local stability of the FM state, when the application of the magnetic force
theorem is strictly justified and the second derivative of the total
energy can be rigorously expressed through the change of the KS
single-particle energies. Fig.~\ref{fig.SFMOsw} shows that it may correct
to a certain extent the values of the second derivatives at the
Brillouin zone boundary. However, it does not change the principal conclusion
and the FM state remains unstable in the large portion of the
Brillouin zone~\cite{PRB02}.\footnote{
Note that in order to prove that the magnetic state is unstable, it is
sufficient to find at least one configuration of the spin-magnetization
density for which $\frac{d^2}{d \theta^2} E({\bf q},\theta)$$<$$0$.}

  2. The second derivatives of the total energy can be mapped onto the
Heisenberg model
$\frac{d^2}{d \theta^2} \left. E({\bf q},\theta) \right|_{\theta = 0}$$=$$J_0$$-$$J_{\bf q}$.
The mapping gives us parameters of magnetic interactions in the reciprocal space
$\{ J_{\bf q} \}$, which can be further transformed to the real space.
The real-space parameters, listed in Table~\ref{tab:JSFMO}, provide a complementary
piece of information which allows to rationalize results of frozen
spin wave calculations and present them in terms of
competition between FM NN and AFM next-NN
interactions in the Fe sublattice (correspondingly $J_1$ and $J_2$ in
Table~\ref{tab:JSFMO}).
\begin{table}[h!]
\caption{Parameters of magnetic interactions in the real space obtained in the
(relaxed) frozen spin wave (SW) and Green's function (GF) calculations near
the FM and AFM
states of Sr$_2$FeMoO$_6$ (in meV). Two values of $J_1$ in the case of the AFM
alignment correspond to nearest-neighbor interactions between Fe sites
having the same (the first number) and opposite (the second number) spins in the
type-II antiferromagnetic structure.}
\label{tab:JSFMO}
\begin{center}
\begin{tabular}{cccc}\hline
method      & $J_1$      & $J_2$     &   $J_{\rm Fe-Mo}$        \\\hline
SW (FM)     &  9.3       & -26.9     &     $-$                  \\
GF (FM)     &  6.2       & -21.5     &     -1.2                 \\
SW (AFM)    & 16.4, 18.6 & -15.2     &       0                  \\\hline
\end{tabular}
\end{center}
\end{table}
The AFM interaction $J_2$ clearly dominates and readily explains why
the FM state becomes unstable.\footnote{Note also that
$J_1$ and
$J_2$ have different coordination numbers (correspondingly 12 and 6).}
An additional piece of information can be obtained from Green's
function calculations, which allow to estimate the direct Fe-Mo
interaction $J_{\rm Fe-Mo}$. However, this interaction is not particularly strong and
does not alter the main conclusion about stability of the FM state.\footnote{
Small difference of the parameters obtained in the frozen spin wave and
Green's function approaches is related
with the fact that $J_{\rm Fe-Mo}$ is already included in the definition
of $J_1$ and $J_2$ in the frozen spin wave calculations.}

  3. If the FM state is unstable, what is the true magnetic ground state of the
ordered Sr$_2$FeMoO$_6$? The question is not simple, because the AFM phase
appears to be also unstable, as it is expected for small concentrations
of ($t_{2g}$) carriers
moving in the AFM background of localized ($e_g$) spins and interacting with
the latter via the strong intra-atomic Hund's coupling~\cite{deGennes}.
This fact is also related with the magnetic-state dependence of interatomic
magnetic interactions, which leads to the
reversed inequality $J_1$$>$$|J_2|$ for parameters calculated in the
AFM state (Table~\ref{tab:JSFMO}). Therefore, the true ground state of the chemically
ordered Sr$_2$FeMoO$_6$
should lie in-between the FM and AFM states. The character of
interatomic magnetic interactions suggests that probably it is a
spin-spiral with ${\bf q}$$||$$[111]$.

  The FM ordering can be stabilized by the anti-site defects or the local
lattice distortions~\cite{PRB02}. However, both mechanisms will destroy
the half-metallic character of the electronic structure of Sr$_2$FeMoO$_6$.
This is an irony of the situation. From the electronic structure point of view,
the most efficient way to stabilize the ferromagnetism is to create holes in
the $\uparrow$-spin Fe($e_g$) band,
by shifting the Fermi level to the lower energy part of the spectrum,
and thereby activate a very effective
channel for the FM DE interactions associated with the
$e_g$ states, similar to the CMR manganites.

  Let us give a rough estimate for this effect.
The width of the Fe($e_g$) band ($W$) in
Sr$_2$FeMoO$_6$ is about 2 eV (Fig.~\ref{fig.SFMOdos}),
which is approximately
two times smaller than the typical $e_g$ bandwidth in FM manganites.
Therefore, the DE and SE interactions in Sr$_2$FeMoO$_6$,
being proportional to $W$ and $W^2$,
will be
correspondingly two and four times smaller
in comparison with the same interactions in manganites.
The latter can be estimated as
$J^D \simeq 80$ and $J^S \simeq - 70$ meV, respectively~\cite{Springer}.
Therefore, the partial depopulation of the $\uparrow$-spin Fe($e_g$) band
in Sr$_2$FeMoO$_6$ leads the following estimate for next-nearest-neighbor
coupling:
$J_2 \simeq \frac{1}{2}J^D + \frac{1}{4}J^S \simeq 20$ meV, which
easily explains the
appearance of ferromagnetism in this compound.

  Thus, the half-metallic electronic structure realized in the FM state of the
ordered double perovskites is a spurious effect since it is largely incompatible with
stability of this FM state. It also implies that the 100\%
ordered double perovskites cannot be ferromagnetic. The latter observation is
qualitatively consistent with the behavior of Sr$_2$FeWO$_6$, where almost perfect ordering of
Fe and W coexists with an AFM ground state~\cite{KobayashiJMMM}.\footnote{
The assignment is based on the analysis of magnetization data, which cannot
exclude the non-collinear spin-spiral alignment.}

  The quantitative theory of ferromagnetism in Sr$_2$FeMoO$_6$ and Sr$_2$FeReO$_6$
is still missing. The exceptionally high value of the
Curie temperature is one of the
main puzzles, which is probably related with the extrinsic
inhomogeneities (anti-site defects, local lattice distortions, grain boundaries)
existing in the samples. The magneto-resistive behavior is based on the
percolative scenario,
and is not necessary related with the half-metallic character of the
electronic structure.
The main difference from CMR manganites is that neither FM nor AFM states are
locally stable and cannot be the total energy minima in the case of double
perovskites. Therefore, the extrinsic inhomogeneity seems to be indispensable
in order to stabilize the FM islands and form the two-pase state similar
to the one depicted in Fig.~\ref{fig.percolation}.\footnote{
This was also supported by experimental studies of Sr$_2$FeW$_{1-x}$Mo$_x$O$_6$
alloy, which shows clear tendency to the phase segregation~\cite{KobayashiJMMM}.
The largest magneto-resistance was observed around $x$$=$$0.15$.}
Then, the negative magneto-resistance may occur if the size of these islands
can be easily controlled by the external magnetic field.

  Finally, our analysis was based on the electronic structure obtained in
LSDA, which may be imperfect. The role of Coulomb correlations on the top
of LSDA is rather unclear. It is also unclear whether the
Coulomb correlations alone can stabilize the FM state. On the one hand, the
on-site $U$ enters the denominator for the SE interactions and therefore suppress the
AFM contribution associated with the Fe($e_g$) states. On the other hand, it
may change the relative position of the Fe and Mo states and thereby
suppress the FM DE interactions associated with the
hybrid $t_{2g}$ band~\cite{PRB02}. We expect that the LSDA band structure
depicts the main problems of the double perovskites, though
the concrete details may depend on the Coulomb correlations on the top
of this picture.

\subsection{$A_2$Mo$_2$O$_7$ ($A$$=$ Y, Gd, and Nd) Pyrochlores}
\label{SubSec.pyrochlores}

\begin{figure}[h!]
\centering \noindent
\resizebox{10cm}{!}{\includegraphics{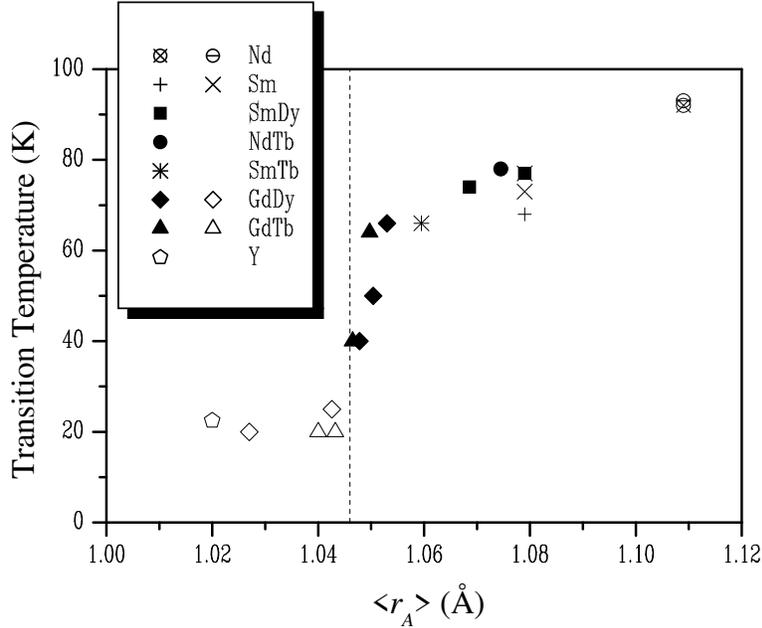}}
\caption{Experimental phase diagram of $A_2$Mo$_2$O$_7$: transition
temperature against the averaged ionic radius. The dotted line shows the
boundary between normal ferromagnetic (right) and spin-glass (left)
behavior.}
\label{fig.pyphase}
\end{figure}

  Pyrochlores with the chemical formula $A_2$Mo$_2$O$_7$ ($A$ being the
divalent element) is one of the rare examples of magnetic oxides on the basis of
$4d$ elements.\footnote{Another famous example of magnetic $4d$
oxides is the ruthenates, crystallizing in the cubic, layered, and double-layered
perovskite structures~\cite{Singh_Springer}.}
The magnetic phase diagram of $A_2$Mo$_2$O$_7$ is controlled by the
averaged ionic radius $\langle r_A \rangle$ of the $A$-sites.
This dependence is very puzzling
(Fig.~\ref{fig.pyphase})~\cite{Katsufuji,Moritomo,Taguchi,Iikubo}.
Large $\langle r_A \rangle$ ($>$$R_c$$\sim$$1.047$) stabilizes the ferromagnetic (FM)
ground state. Typical examples of the FM pyrochlores are Nd$_2$Mo$_2$O$_7$ and
Gd$_2$Mo$_2$O$_7$. The Curie temperature ($T_C$) is of the order of 80 K and slowly increases
with $\langle r_A \rangle$. Smaller $\langle r_A \rangle$ ($<$$R_c$) give rise to
the spin-glass (SG) behavior. The characteristic transition temperature to the SG state
is of the order of 20 K.
The canonical example of the SG compounds is Y$_2$Mo$_2$O$_7$~\cite{Y2Mo2O7_basicExp}.
Intuitively, it is clear that such a behavior should be related with the change of
the NN exchange coupling, which becomes
antiferromagnetic in the SG region.
The pyrochlore lattice supplemented
with the AFM exchange
interactions presents a typical example of
geometrically frustrated systems with an infinitely degenerate magnetic ground state~\cite{Reimers},
which probably predetermines the SG behavior.\footnote{
Nevertheless, many details of this behavior remain to be understood. For example,
according to the recent experimental data~\cite{Y2Mo2O7_recentExp}, the SG state
appears due to the joint effect of geometrical frustrations and the
disorder of local lattice distortions. The latter is responsible
for the spacial
modulations
of interatomic magnetic interactions, which
freeze the random spin configuration.}
The FM-SG transition in Mo pyrochlores is closely related with the metal-insulator transition.
All compounds which reveal the SG behavior are
small-gap insulators. However, the opposite statement is generally incorrect and in
different compounds the ferromagnetism is known to coexist with the metallic as well as
the insulating behavior~\cite{Iikubo}.
  In this section we will try to understand which parameter
of the crystal structure controls the sign of the NN magnetic interactions, and which
part of the electronic structure is responsible for the FM and AFM interactions
in these compounds.

\subsubsection{Main Details of Crystal and Electronic Structure}
\label{SubSubSec.str_pyrochlores}

  The pyrochlores $A_2$Mo$_2$O$_7$
crystallize in a face-centered cubic
structure with the space group $Fm\overline{3}d$, in which $A$ and Mo
occupy correspondingly $16d$ and $16c$ positions, and form interpenetrating
sublattices of corner-sharing tetrahedra. There is
only one internal parameter which may control the properties
of $A_2$Mo$_2$O$_7$. That is the coordinate $u$ of the O $48f$ sites.

\begin{figure}[h!]
\centering \noindent
\resizebox{8cm}{!}{\includegraphics{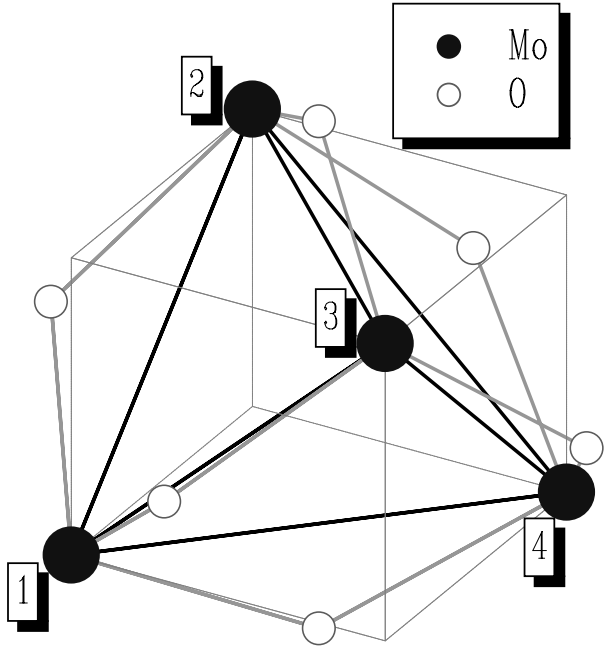}}
\caption{Positions of the Mo and O $48f$ sites in the pyrochlore
lattice.}
\label{fig.pystructure}
\end{figure}

  The single Mo tetrahedron is shown in Fig.~\ref{fig.pystructure}.
Four Mo sites are located at
$\bm{\tau}_1$$=$$(0,0,0)$,
$\bm{\tau}_2$$=$$(0,\frac{1}{4},\frac{1}{4})$,
$\bm{\tau}_3$$=$$(\frac{1}{4},0,\frac{1}{4})$,
and $\bm{\tau}_4$$=$$(\frac{1}{4},\frac{1}{4},0)$,
in units of cubic lattice parameter $a$.
Each Mo site has sixfold O $48f$ coordination. The oxygen atoms specify the local
coordinate frame around each Mo site. Around 1, it is given by
\begin{equation}
{\cal R}_1^{\alpha \beta} = \frac{1+(1-8u)\delta_{\alpha \beta}}{\sqrt{64u^2-32u+6}},
\label{eqn:lframe}
\end{equation}
where $\alpha ,\beta$$=$ $x$, $y$, or $z$.
$u$$=$$\frac{5}{16}$ corresponds to the perfect octahedral environment, while
$u$$>$$\frac{5}{16}$ gives rise to an additional
trigonal contraction of the local coordinate frame. Similar matrices associated with the sites
2, 3, and 4 can be obtained by the $180^\circ$ rotations of
$\widehat{\cal R}_1$ around
$x$, $y$, and $z$, respectively.

\begin{figure}[h!]
\centering \noindent
\resizebox{10cm}{!}{\includegraphics{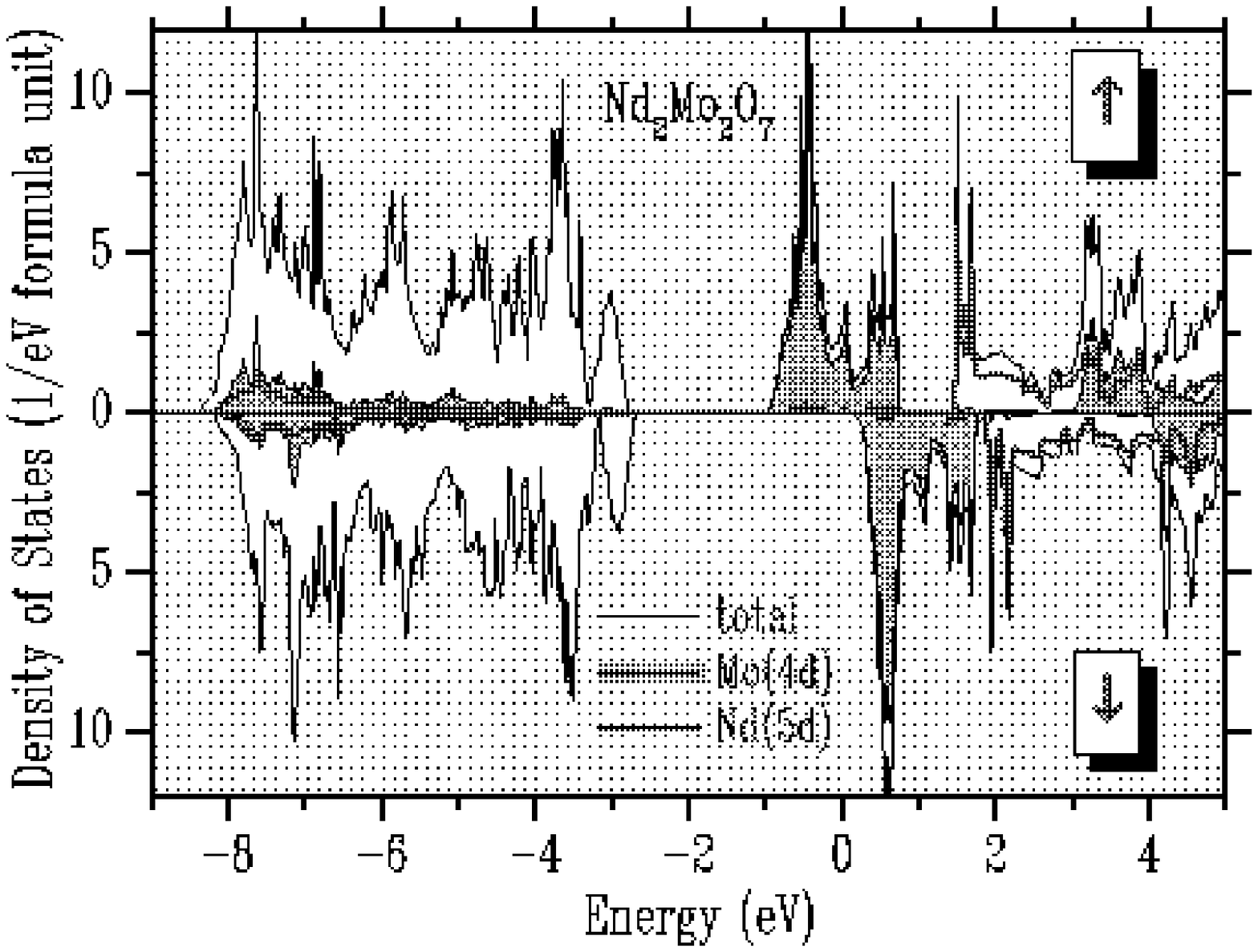}}
\caption{Total and partial densities of states of Nd$_2$Mo$_2$O$_7$
in the local-spin-density approximation. The Mo($t_{2g}$) states are
located near the Fermi level (chosen as zero of energy). The
Mo($e_g$) states emerge around 4 eV.}
\label{fig.pyLSDAdos}
\end{figure}

\begin{figure}[h!]
\centering \noindent
\resizebox{8cm}{!}{\includegraphics{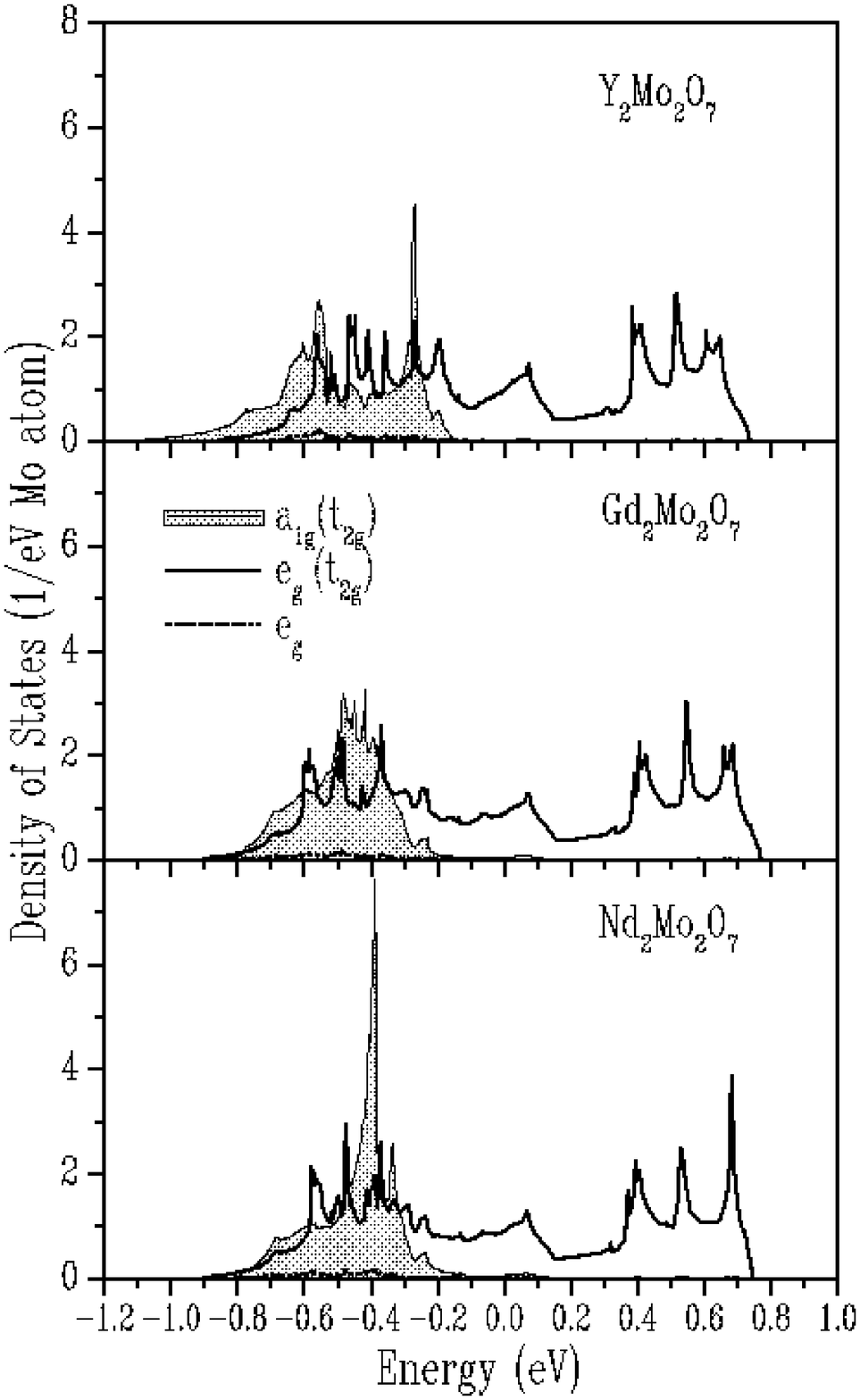}}
\caption{Mo($t_{2g}$) states in the local coordinate frame, split into the
one-dimensional $a_{1g}$ and two-dimensional $e_g'$$=$$e_g$($t_{2g}$)
representations by
the local trigonal distortion.}
\label{fig.pyDOSt2g}
\end{figure}

  Structural parameters of $A_2$Mo$_2$O$_7$
(taken from Refs.~\cite{Katsufuji,Moritomo,Taguchi,Iikubo,Y2Mo2O7_basicExp})
are listed in Table~\ref{tab:pystructure}.
\begin{table}
\caption{Structural parameters of $A_2$Mo$_2$O$_7$ ($A$$=$ Y, Nd, and Gd):
the cubic lattice parameter $a$ (in \AA),
positions of O $48f$ sites $[u,\frac{1}{8},\frac{1}{8}]$ (units of $a$),
the distances Mo-Mo (in \AA), and the angles Mo-O-Mo
(in degrees).}
\label{tab:pystructure}
\begin{center}
\begin{tabular}{ccccc}\hline
compound          &   $a$       &    $u$    & $d_{\rm Mo-Mo}$ & $\angle$Mo-O-Mo  \\\hline
Y$_2$Mo$_2$O$_7$  & 10.21       &  0.33821  & 3.6098          & 127.0            \\
Gd$_2$Mo$_2$O$_7$ & 10.3356     &  0.33158  & 3.6542          & 130.4            \\
Nd$_2$Mo$_2$O$_7$ & 10.4836     &  0.32977  & 3.7065          & 131.4            \\\hline
\end{tabular}
\end{center}
\end{table}
Corresponding densities of states, obtained in
LSDA are shown in Figs.~\ref{fig.pyLSDAdos} and \ref{fig.pyDOSt2g}.

  In the local coordinate frame, the Mo($4d$) orbitals are split into the triply-degenerate
$t_{2g}$ and doubly-degenerate $e_g$ states. Twelve $t_{2g}$ bands
are located near the
Fermi level and well separated from the rest of the spectrum.
Interestingly enough that
all three compounds $A_2$Mo$_2$O$_7$ ($A$$=$ Y, Gd, and Nd)
are ferromagnetic, even in
LSDA, that is rather unusual for the $4d$ oxides, perhaps except the well know
example of SrRuO$_3$~\cite{Singh}.
The trigonal distortion
and the different hybridization with the O($2p$) states
will further split the Mo($t_{2g}$) states into the
one-dimensional $a_{1g}$ and two-dimensional $e_g'$ representations.\footnote{
In the local coordinate frame, the $a_{1g}$ and two $e_g'$ orbitals
have the following form: $|a_{1g}\rangle$$=$$\frac{1}{\sqrt{3}}
(|xy\rangle$$+$$|yz\rangle$$+$$|zx\rangle)$,
$|e_{g1}'\rangle$$=$$\frac{1}{\sqrt{2}}(|yz\rangle$$-$$|zx\rangle)$, and
$|e_{g2}'\rangle$$=$$\frac{1}{\sqrt{6}}($$-$$2|xy\rangle$$+$$|yz\rangle$$+$$
|zx\rangle)$.}
The crystal structure affects the Mo($t_{2g}$) bandwidth via two mechanisms
(see Table~\ref{tab:pystructure}).
\begin{enumerate}
\item
The Mo-O-Mo angle, which decreases in the direction Nd$\rightarrow$Gd$\rightarrow$Y.
Therefore,
the interactions between Mo($t_{2g}$) orbitals which are mediated by the
O($2p$) states will also decrease.
\item
The lattice parameter $a$ and the Mo-Mo distance, which decrease in the
direction Nd$\rightarrow$Gd$\rightarrow$Y by 2.6\%.
This will increase the direct
interactions between extended Mo($4d$) orbitals.
\end{enumerate}
Generally, these two effects act in the opposite directions and partly
compensate each other. For example, the width of the $e_g'$ band is practically
the same for all three compounds (Fig.~\ref{fig.pyDOSt2g}). On the other hand, the
$a_{1g}$ orbitals, whose lobes are the most distant from all neighboring oxygen
sites are mainly affected by the second mechanism, and the $a_{1g}$
bandwidth will {\it increase} in the direction Nd$\rightarrow$Gd$\rightarrow$Y.
As we will see below, this effect plays a crucial role
in the stabilization of AFM interactions in Y$_2$Mo$_2$O$_7$,
which predetermines SG behavior.

  Thus, despite an apparent complexity of the crystal structure,
the pyrochlores $A_2$Mo$_2$O$_7$ present a rather simple example of the
electronic structure and in order to understand
the nature of fascinating
electronic and magnetic properties of these compounds,
we need to concentrate on the behavior of twelve well isolated
Mo($t_{2g}$) bands.\footnote{In this sense, the physics is similar to the spinel
compounds~\cite{Anisimov}.
Note, however, that the oxygen coordination is
very different in the spinel and pyrochlore structures.}

  In this section we consider results of model Hartree-Fock
calculations~\cite{PRB03.2}, which combine
fine details of the electronic structure for these bands,
extracted from LSDA, and
the on-site Coulomb interactions among the Mo($4d$) electrons, treated
in the most general
rotationally invariant form~\cite{PRL98}.\footnote{
The on-site Coulomb interactions are specified by three radial
Slater integrals, $F^0$, $F^0$, and $F^0$.
Equivalently, one can introduce parameters
of the averaged Coulomb interaction, $U$$=$$F^0$,
the intra-atomic exchange coupling, $J$$=$$\frac{1}{14}(F^2$$+$$F^4)$, and the
non-sphericity of Coulomb interactions between orbitals with the
same spin, $B$$=$$\frac{1}{441}(9F^2$$-$$5F^4)$.
$J$$\simeq$$0.5$ eV is taken from the constraint-LSDA calculations~\cite{PRB94}.
$B$ can be
estimated from $J$ using the ratio $F^4/F^2$$\simeq$$0.63$, which
holds for the Slater integrals in the atomic limit.
This yields
$B$$\simeq$$0.06$ eV. The Coulomb $U$ is treated as the parameter
in order to consider different
scenarios, covering both metallic and insulating behavior of
$A_2$Mo$_2$O$_7$. The constraint-LSDA calculations for Mo compounds yield
$U$$\approx$$3.0$ eV~\cite{PRB94}. This value can be further reduced by allowing for
the (proper) $e_g$ electrons to participate in the screening of on-site Coulomb
interactions of the $t_{2g}$ electrons~\cite{PRB96}.}

\subsubsection{Effects of Crystal Structure and On-Site Coulomb Interactions
on the Interatomic Exchange Coupling}
\label{SubSubSec.exchange_pyrochlores}

\begin{figure}[h!]
\centering \noindent
\resizebox{8cm}{!}{\includegraphics{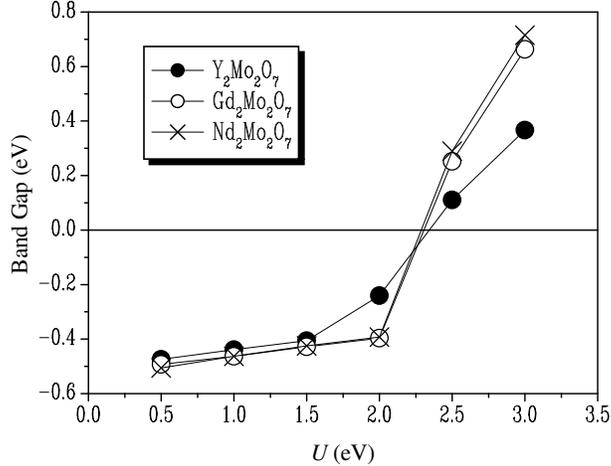}}
\caption{The band gap
as a function of Coulomb $U$. For
$U$$\leq$$2.0$ eV there is an overlap between bands, corresponding
to the negative value of the band gap, while
$U$$\geq$$2.5$ eV opens the real gap.}
\label{fig.gap}
\end{figure}

\begin{figure}
\centering \noindent
\resizebox{8cm}{!}{\includegraphics{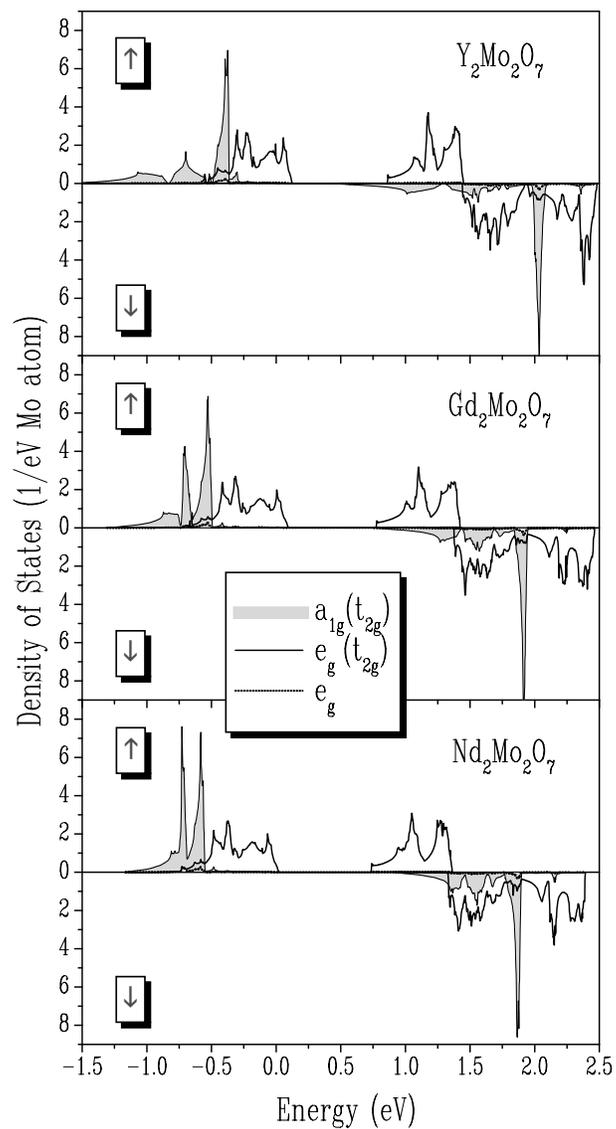}}
\caption{Distribution of the Mo($4d$) states
obtained in model Hartree-Fock
calculations for $U$$=$$3.0$ eV.}
\label{fig.pyDOS30}
\end{figure}

  According to the LSDA calculations for the FM state (Fig.~\ref{fig.pyDOSt2g}), the
$\uparrow$-spin $a_{1g}$ band is fully occupied and the Fermi level crosses
the doubly-degenerate $e_g'$ band. Therefore, it is clear that at some point the
Coulomb $U$ will split the $e_g'$ band and form an insulating state with
the spontaneously
broken $Fm\overline{3}d$ symmetry. Such situation occurs between $U$$=$ 2.0 and 2.5 eV
for all considered compounds (Fig.~\ref{fig.gap}).
Typical densities of states in the insulating phase are shown in Fig.~\ref{fig.pyDOS30}.

\begin{figure}[h!]
\centering \noindent
\resizebox{6cm}{!}{\includegraphics{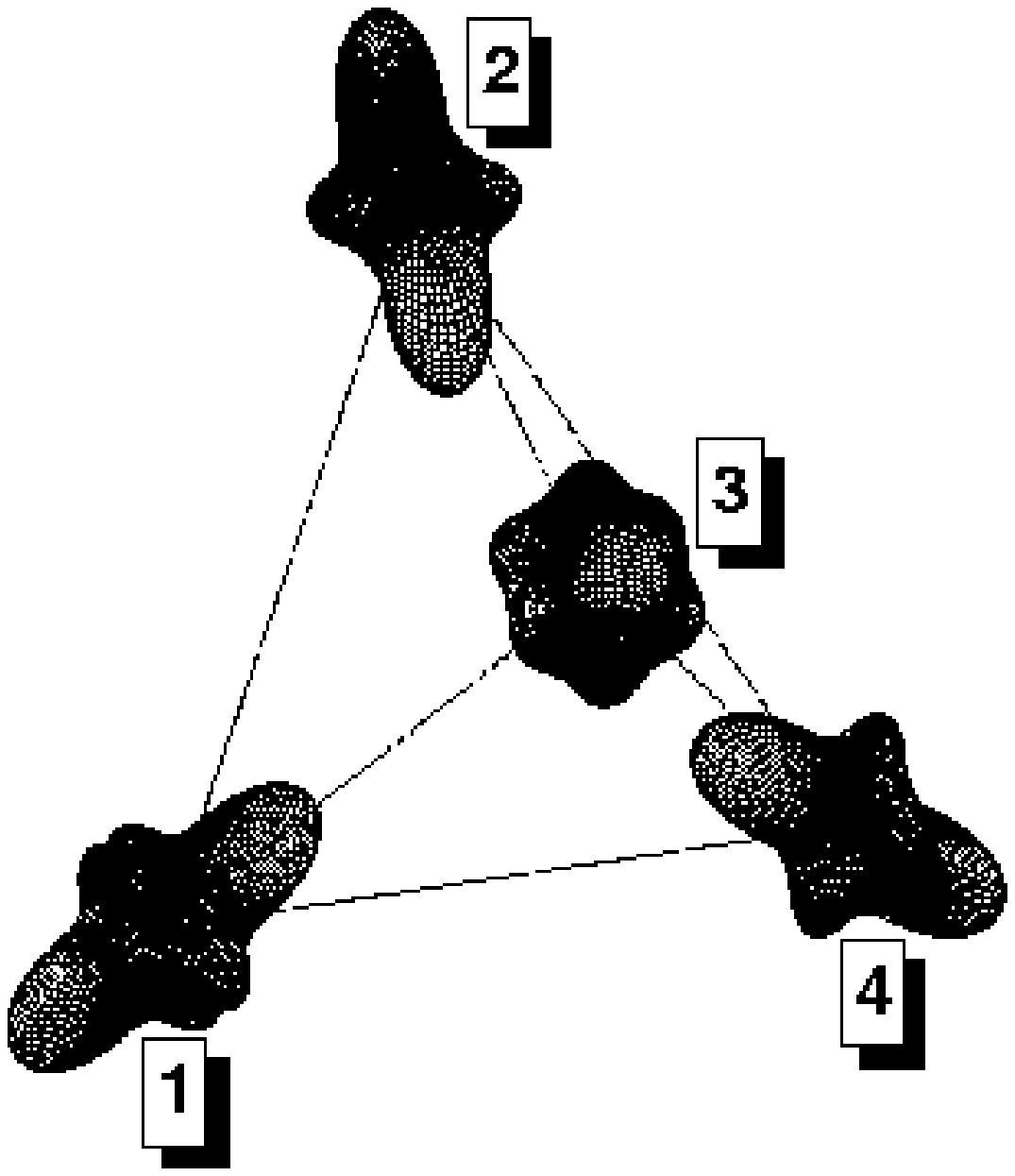}}
\caption{Orbital ordering in Nd$_2$Mo$_2$O$_7$ for $U$$=$$1.5$ eV.}
\label{fig.pyOO15}
\end{figure}

\begin{figure}[h!]
\centering \noindent
\resizebox{6cm}{!}{\includegraphics{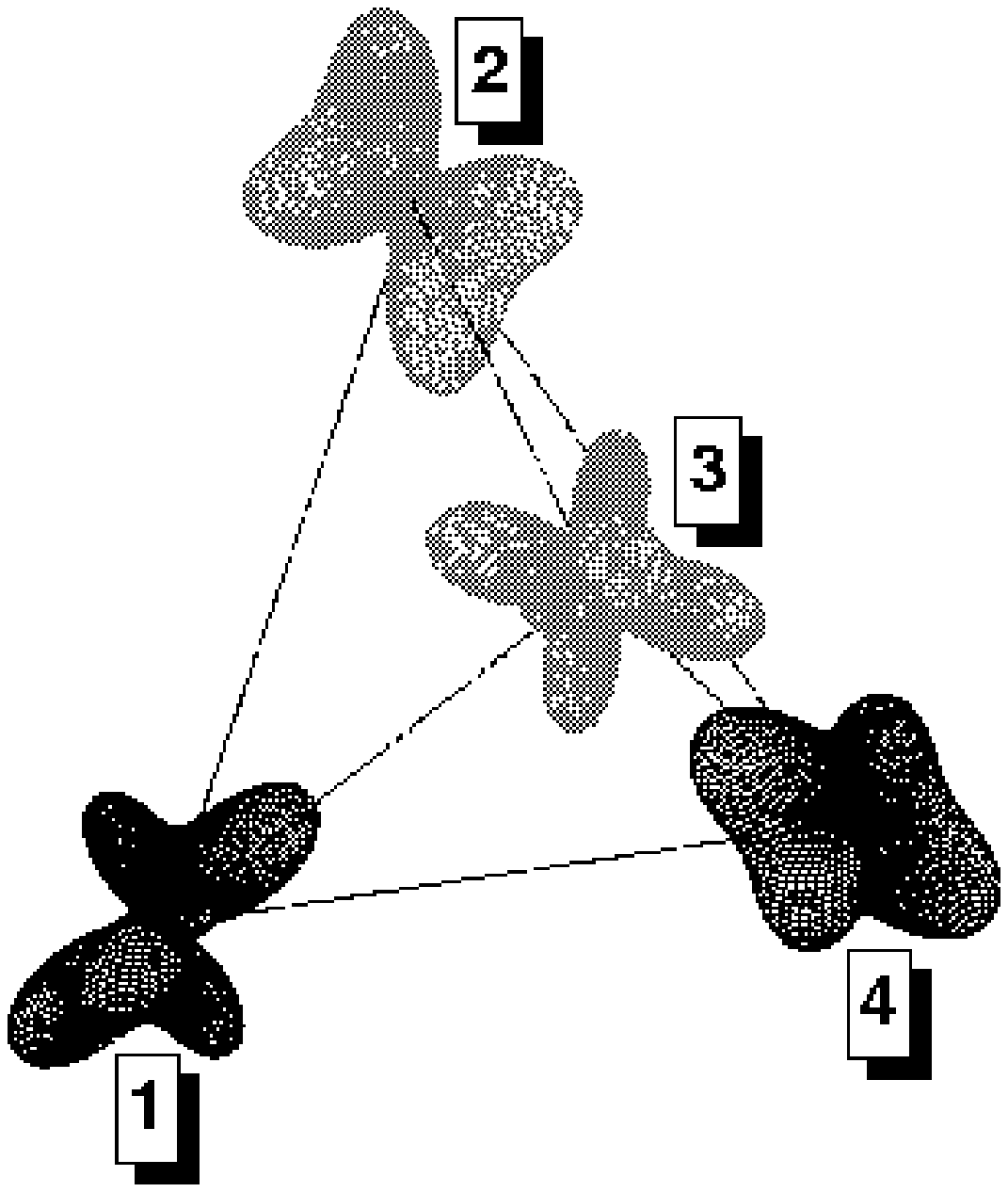}} \resizebox{6cm}{!}{\includegraphics{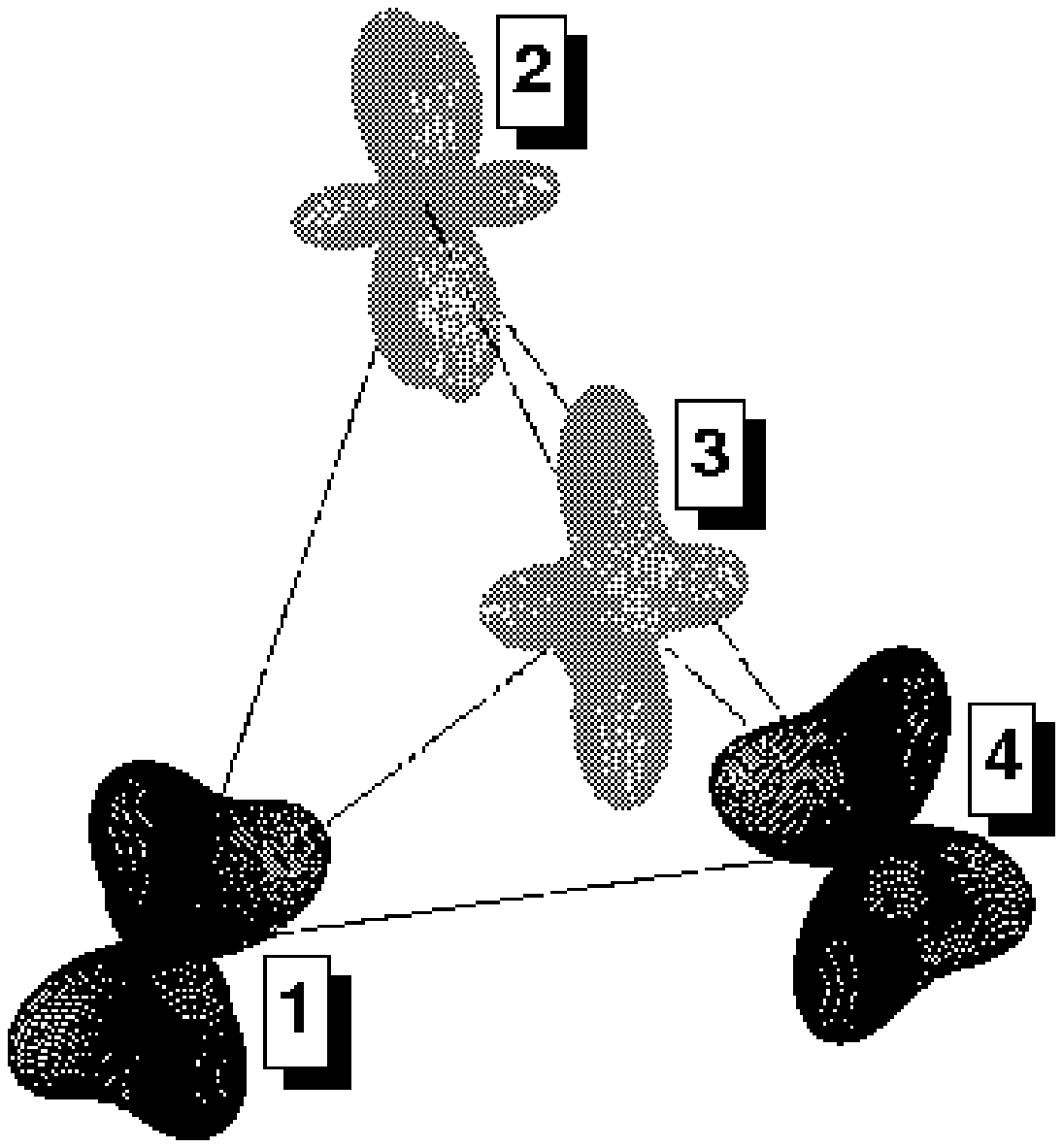}}
\caption{Orbital ordering in Y$_2$Mo$_2$O$_7$ for $U$$=$$3.0$ eV.
Left panel: ferromagnetic phase. Two orbital sublattices are shown by black and gray colors.
Right panel: antiferromagnetic phase. Two spin sublattices are show by black and gray colors.}
\label{fig.pyOO30}
\end{figure}

  The insulating behavior is accompanied by the change of
the orbital ordering (the distribution of the Mo($4d$) electron
densities). In the metallic state, it
comes exclusively from the local trigonal distortions of the oxygen octahedra and
represents the alternating $a_{1g}$ orbital densities in the background of degenerate
$e_g'$ orbitals (Fig.~\ref{fig.pyOO15}).
In the insulating state, the orbital ordering
is determined not only by the local trigonal distortions, but also
by the form of SE interactions between NN Mo sites,
and tends to minimize
the energy of these interactions~\cite{KugelKhomskii}.
Two typical examples for the FM
and AFM (obtained after the flip of magnetic moments at the sites 2 and 3) phases
are shown in Fig.~\ref{fig.pyOO30}.
As expected for the FM spin ordering~\cite{KugelKhomskii},
the $e_g'$ orbitals tend to order ''antiferromagnetically'' and
form two Mo sublattices.
Clearly, this orbital
ordering breaks the $Fm\overline{3}d$ symmetry: if the sites belonging to the
same sublattice can still be transformed to each other using the symmetry operations
of the $Fm\overline{3}d$ group, the sites belonging to different sublattices -- cannot.
This will generally lead to the anisotropy of electronic properties, including the
NN magnetic interactions.
The AFM spin ordering coexists with the FM orbital ordering.
It breaks the $Fm\overline{3}d$ symmetry in the spin sector, but not
in the orbitals one.

\begin{figure}[h!]
\centering \noindent
\resizebox{8cm}{!}{\includegraphics{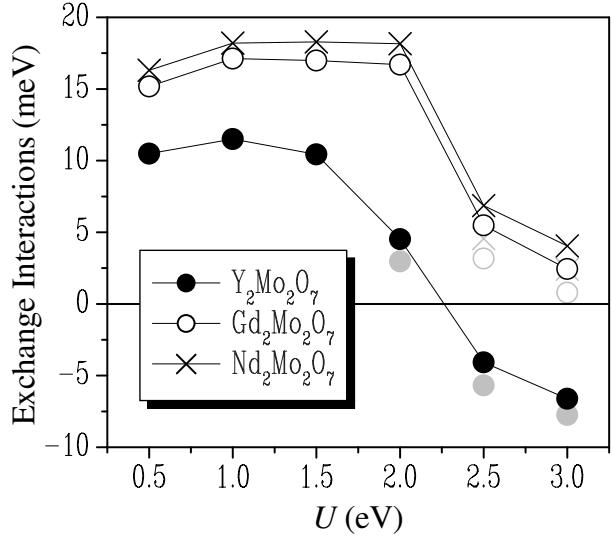}}
\caption{Nearest-neighbor exchange interactions calculated in the
ferromagnetic state. The orbital ordering realized for large
$U$ breaks the $Fm\overline{3}d$ symmetry, that
leads to the inequality
$J_{12}$$=$$J_{13}$$\neq$$J_{14}$. Two such parameters, $J_{12}$
and $J_{14}$, are shown by dark and light symbols, respectively.}
\label{fig.pyexchange}
\end{figure}

  Results for NN magnetic interactions are
shown in Fig.~\ref{fig.pyexchange}, as a function of $U$.
We note the following.
\begin{enumerate}
\item
$J_{\bm{\tau} \bm{\tau}'}$, which are ferromagnetic for small $U$, exhibit a sharp
drop at the point of transition into the insulating state.
\item
There is a significant difference between Nd/Gd and Y based compounds: in the case of Y,
the exchange parameters are almost rigidly shifted towards negative values, so that
the NN coupling become
antiferromagnetic in the insulating phase, while it
remains ferromagnetic in the case of Nd and Gd.
\end{enumerate}

\begin{figure}[h!]
\centering \noindent
\resizebox{8cm}{!}{\includegraphics{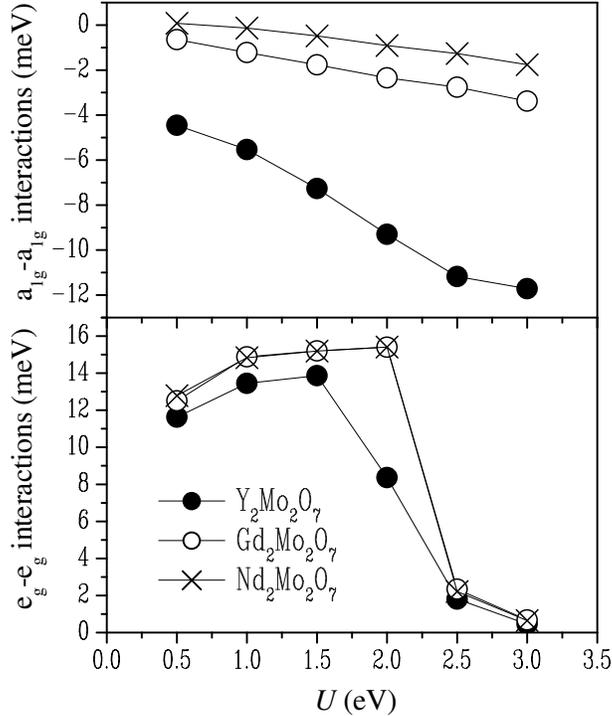}}
\caption{Contributions of $a_{1g}$ and $e_g'$
orbitals (coming from the $t_{2g}$ manifold in the local
coordinate frame) to the exchange constant $J_{14}$.}
\label{fig.pyjpartial}
\end{figure}

    The behavior can be easily understood by considering partial $a_{1g}$ and
$e_g'$ contributions to the NN exchange coupling, calculated after
transformation to the local coordinate frame at each site of the system
(Fig.~\ref{fig.pyjpartial}).
Large FM $e_g'$-$e_g'$ interaction in the metallic regime is related with
the double exchange (DE) mechanism, which is the measure of the kinetic
energy for the itinerant $\uparrow$-spin $e_g'$ electrons.
As long as the system is metallic, the DE interactions are not
sensitive to the value of $U$, and the FM coupling will prevail.
In the insulating state,
the $e_g'$ electrons are localized
at the atomic orbitals. This reduces the kinetic energy
and suppresses
the DE interactions, that explain the sharp drop of $J_{\bm{\tau} \bm{\tau}'}$
(Fig.~\ref{fig.pyexchange}).

  The main difference between Y and Nd/Gd based compounds is related with the
$a_{1g}$-$a_{1g}$ interaction. Since the $\uparrow$-spin $a_{1g}$ band is
fully occupied and the $\downarrow$-spin band is empty, this interaction is
antiferromagnetic due to the superexchange mechanism.
Since the SE coupling is proportional to the square of the $a_{1g}$ bandwidth,
this interaction will be the largest in the case of Y, that explains the AFM
character of the total exchange
coupling realized in this compound for large $U$.

  This example clearly shows at the importance of Coulomb $U$
in the problem interatomic magnetic interactions of Mott insulators:
it is absolutely
indispensable in order to open the gap between occupied and unoccupied states,
and suppress the FM DE interactions. Only in this case the total coupling between
neighboring Mo sites can become antiferromagnetic, which is a necessary precondition
for the SG behavior. On the contrary, the metallic state of Mo pyrochlores will
always coexist with the ferromagnetism. Therefore, it is not right to say that the
energy gap is not a ground-state property and therefore does not need to
be present
in the Kohn-Sham quasi-particle spectrum of the SDFT. The present example shows exactly
the opposite: the energy gap may determine not only the value, but also the sign
of interatomic magnetic interactions, which are the ground-state properties.

\section{Concluding Remarks}
\label{Sec.Summary}

  In this review we tried to formulate some framework for the analysis of
interatomic magnetic interactions in the transition-metal oxides and give some hints
and ideas illustrating how this analysis should be generally done or at least
started with on the basis of realistic electronic structure calculations.
It is based on the fundamental magnetic force theorem, which allows to connect
the problem of interatomic magnetic interactions in solids with the (generally
auxiliary) single-particle spectrum obtained from the solution of Kohn-Sham equations
in the spin-density-functional theory for the ground state.

 The straightforward application of this theorem is seriously hampered by the
fact that in realistic calculations, SDFT is supplemented with some approximations,
which are not always adequate for TMO. Nevertheless,
in many cases the electronic structure calculations can provide the basic idea or at least
some useful input information for the analysis of interatomic magnetic interactions
in TMO, which (perhaps with some imaginations!) can be transformed to a realistic
semi-quantitative model. This point of view was illustrated for a number of popular
nowadays compounds, such as CMR manganites, double perovskites Sr$_2$Fe$M$O$_6$ ($M$$=$ Mo, Re),
and pyrochlores $A_2$Mo$_2$O$_7$ ($A$$=$ Y, Gd, and Nd).

  It would be unfair to say that this strategy is universal and can be directly applied
to all possible types of TMO. In some sense, we were lucky that the realistic information about
the electronic structure of the considered compounds can be obtained already on the
level of LSDA:
\begin{itemize}
\item
It gives a rough idea about the competition of double exchange and superexchange
interactions in CMR manganites and double perovskites.
\item
It allows to identify the electronic states, which are mainly responsible for the
physics of Mo-pyrochlores and well separated from the rest of the spectrum, so that
the next step connecting the LSDA calculations with the degenerate Hubbard model
becomes basically a matter of routine.
\end{itemize}

  Unfortunately, such a generalization is not always possible and there is a wide
class of magnetic TMO, which still presents a very challenging
problem for the
first-principles electronic structure calculations. This group
includes many nickelates and cuprates. These systems are especially complicated
because of two reasons:
\begin{itemize}
\item
For many of them, the standard LSDA calculations erroneously lead to the nonmagnetic
ground state (the most famous example is La$_2$CuO$_4$~\cite{Pickett}).
Therefore, it is difficult to
make any conjectures basing on the LSDA.
\item
All these compounds are of the charge-transfer type. In this case, the semi-empirical
LDA$+$$U$ calculations may not be very reliable either.
\end{itemize}
In order to proceed with the realistic description of electronic and
magnetic properties of these materials, the development of new methods of electronic
structure calculations, which would go beyond LSDA and get rid of several empirical
features of the LDA$+$$U$ approach is a very important task.

\end{sloppypar}

\newpage

\begin{thebibliography}{10}

\bibitem{Goodenough}
Goodenough, J.B. 1955,
Phys.~Rev. 100, 564.

\bibitem{AndersonHasegawa}
Anderson, P.W. and Hasegawa, H. 1955,
Phys.~Rev. 100, 675.

\bibitem{Kanamori}
Kanamori, J. 1959,
J.~Phys.~Chem.~Solids~10, 87.

\bibitem{Anderson}
Anderson, P.W. 1959,
Phys.~Rev.~115, 2.

\bibitem{Oguchi}
Oguchi, T., Terakura, K., and Williams, A.R. 1983,
Phys.~Rev.~B 28, 6443.

\bibitem{Terakura}
Terakura, K., Oguchi, T., Williams, A.R., and K\"ubler, J. 1984,
Phys.~Rev.~B 30, 4734.

\bibitem{deGennes}
De~Gennes, P.-G. 1960,
Phys.~Rev. 118, 141.

\bibitem{KugelKhomskii}
Kugel, K.~I. and Khomskii, D.~I. 1982,
Usp.~Fiz.~Nauk~136, 621.

\bibitem{SDFT}
Hohenberg, P., and Kohn, W. 1964, Phys.~Rev. 136, B864;
Kohn, W., and Sham, L.J. 1965, Phys.~Rev. 140, A1133;
von Bath, U., and Hedin, L. 1972, J.~Phys.~C: Solid States Phys. 5, 1629.

\bibitem{Stoner}
Gunnarsson, O. 1976, J.~Phys.~F: Metal Phys. 6, 587;
Janak, J.F. 1977, Phys.~Rev.~B 16, 255.

\bibitem{AZA}
Anisimov, V.I., Zaanen, J., and Andersen, O.K. 1991,
Phys.~Rev.~B 44, 943.

\bibitem{PRB94}
Solovyev, I.V., Dederichs, P.H., and Anisimov, V.I. 1994,
Phys.~Rev.~B 50, 16861.

\bibitem{PRL98}
Solovyev, I.V., Liechtenstein, A.I., and Terakura, K. 1998,
Phys.~Rev.~Lett. 80, 5758.

\bibitem{LMTO}
Andersen, O.K. 1975, Phys.~Rev.~B 12, 3060;
Gunnarsson, O., Jepsen, O., and Andersen, O.K. 1983,
Phys.~Rev.~B 27, 7144;
http://www.mpi-stuttgart.mpg.de/andersen/.

\bibitem{OEP}
Talman, J.D., and Shadwick, W.F. 1976,
Phys.~Rev.~A 14, 36.

\bibitem{Kotani}
Kotani, T., and Akai, H. 1996,
Phys.~Rev.~B 54, 16502;
Kotani, T. 1998,
J.~Phys.: Condens. Matter 10, 9241.

\bibitem{GW_review}
Aryasetiawan, F., and Gunnarsson, O. 1998,
Rep.~Prog.~Phys. 61, 237.

\bibitem{PRB98}
Solovyev, I.V., and Terakura, K. 1998,
Phys.~Rev.~B 58, 15496.

\bibitem{VignaleRasolt}
Vignale, G., and Rasolt, M. 1987,
Phys.~Rev.~Lett. 59, 2360;
Vignale, G., and Rasolt, M. 1988,
Phys.~Rev.~B 37, 10685;
Rasolt, M., and Vignale, G. 1990,
Phys.~Rev.~Lett. 65, 1498.

\bibitem{Bruno}
Bruno, P. 2003,
Phys.~Rev.~Lett. 90, 087205.

\bibitem{Liechtenstein}
Liechtenstein, A.I., Katsnelson, M.I., Antropov, V.P., and
Gubanov, V.A. 1987, J.~Magn.~Magn.~Matter. 67, 65.

\bibitem{intranc}
Nordstr\"{o}m, L., and Singh, D.J. 1996,
Phys.~Rev.~Lett. 76, 4420;
Oda, T., Pasquarello, A., and Car, R. 1998,
Phys.~Rev.~Lett. 80, 3622;
Bylander, D.M., Niu, Q., and Kleinman L. 2000,
Phys.~Rev.~B 61, R11875.

\bibitem{Halilov}
Halilov, S.V., Eschrig, H., Perlov, A.Y., and Oppeneer, P.M. 1998,
Phys.~Rev.~B 58, 293.

\bibitem{gBloch}
Herring, C. 1966, in \textit{Magnetism}, Vol.~IV, G.~Rado and H.~Suhl (Ed.),
Academic Press, New York, London;
Sandratskii, L.M. 1986,
Phys.~Status~Sol.~B 136, 167.

\bibitem{PRL99}
Solovyev, I.V., and Terakura, K. 1999,
Phys.~Rev.~Lett. 82, 2959.

\bibitem{Springer}
Solovyev, I.V., and Terakura, K. 2003, in
{\it Electronic Structure and Magnetism of Complex Materials},
D.~J.~Singh and D.~A.~Papaconstantopoulos (Eds.), Springer, Berlin.

\bibitem{SEoxygen}
Zaanen, J., and Sawatzky, G.A. 1987,
Can.~J.~Phys. 65, 1262.

\bibitem{RKKY}
Blackman, J.A., and Elliott, R.J. 1969,
J.~Phys.~C: Solid States Phys. 2, 1670;
Bulaevskii, L.N, and Panyukov, S.V. 1986,
JETP Lett. 43, 240;
Lang, P., Nordstr\"{o}m, L., Wildberger, K., Zeller, R.,
Dederichs, P.H., and Hoshino, T. 1996,
Phys.~Rev.~B 53, 9092;
Levy, P.M., Maekawa, S., and Bruno, P. 1998,
Phys.~Rev.~B 58, 5588.

\bibitem{PRB99}
Solovyev, I.V. 1999,
Phys.~Rev.~B 60, 8550.

\bibitem{MnO}
Lines, M.E., and Jones, E.D. 1965,
Phys.~Rev. 139, A1313;
Kohgi, M., Ishikawa, Y., and Endoh, Y. 1972,
Solid~State~Commun. 11, 391;
Pepy G. 1974, J.~Phys.~Chem.~Solids 35, 433.

\bibitem{DMFT}
Held, K., Keller, G., Eyert, V., Vollhardt, D., and Anisimov, V.I. 2001,
Phys.~Rev.~Lett. 86, 5345;
Savrasov, S.Y. and Kotliar, G. 2003,
Phys.~Rev.~Lett. 90, 056401.

\bibitem{NSMO}
Kuwahara, H., Okuda, T., Tomioka, Y., Asamitsu, A., and Tokura, Y. 1998,
in {\it Science and Technology of Magnetic Oxides},
M.~Hundley, J.~Nickel, R.~Ramesh, and Y.~Tokura (Eds.). MRS Symposia
Proceedings No.~494, Materials Research Society, Pittsburg.

\bibitem{Imada}
Imada, M., Fujimori, A., and Tokura, Y. 1998,
Rev.~Mod.~Phys. 70, 1039.

\bibitem{GordonBreach}
Tokura, Y. (Ed.) 2000,
{\it Colossal Magnetoresistive Oxides},
Gordon and Breach Science Publishers, Tokyo.

\bibitem{Dagotto}
Dagotto, E., Hotta, T., and Moreo, A. 2001,
Phys.~Rep. 344, 1.

\bibitem{Priya}
Mahadevan, P., Shanthi, N., and Sarma, D.D. 1997,
J.~Phys.: Condens.~Matter 9, 3129;
Papaconstantopoulos, D.A., and Pickett, W.E. 1998,
Phys.~Rev.~B 57, 12751.

\bibitem{BrinkKhomskii}
van~den~Brink, J., and Khomskii, D. 1999,
Phys.~Rev.~Lett. 82, 1016.

\bibitem{PRB01}
Solovyev, I.V., and Terakura, K. 2001,
Phys.~Rev.~B 63, 174425.

\bibitem{CEpapers}
Solovyev, I.V., and Terakura, K. 1999,
Phys.~Rev.~Lett. 83, 2825;
Solovyev, I.V. 2001,
Phys.~Rev.~B 63, 174406.

\bibitem{Kino}
Kino, H., Aryasetiawan, F., Miyake, T., Solovyev, I., Ohno, T., and Terakura, K. 2003,
Physica~B (in press).

\bibitem{Lee}
Lee, J., Yu, J., and Terakura, K. 1998,
J.~Korean~Phys.~Soc. 33, S55.

\bibitem{Murakami}
Murakami, Y., Kawada, H., Kawata, H., Tanaka, M., Arima, T., Moritomo, Y.,
and Tokura, Y. 1998, Phys.~Rev.~Lett. 80, 1932.

\bibitem{OKATB}
Andersen, O.K., Liechtenstein, A.I., Jepsen, O., and Paulsen, F. 1995,
J.~Phys.~Chem.~Solids 56, 1573.

\bibitem{SlaterKoster}
Slater, J.C., and Koster, G.F. 1954,
Phys. Rev. 94, 1498.

\bibitem{MullerHartmann}
M\"{u}ller-Hartmann, E., and Dagotto, E. 1996,
Phys.~Rev.~B 54, R6819.

\bibitem{PRB03}
Solovyev, I.V. 2003,
Phys.~Rev.~B 57, 014412.

\bibitem{Maezono}
Maezono, R., Ishihara, S., and Nagaosa, N. 1998,
Phys.~Rev.~B 58, 11583.

\bibitem{Fang}
Fang, Z., Solovyev, I.V., and Terakura, K. 2000,
Phys.~Rev.~Lett. 84, 3172.

\bibitem{Shen}
Shen, S.Q. 2001,
Phys.~Rev.~Lett. 86, 5842.

\bibitem{Yunoki}
Yunoki, S., Hotta, T., and Dagotto, E. 2000,
Phys.~Rev.~Lett. 84, 3714.

\bibitem{Hotta}
Hotta, T., Takada, Y., and Koizumi, H. 1998,
Int.~J.~Mod.~Phys.~B 12, 3437.

\bibitem{SpinValve}
Kuwahara, H., Okuda, T., Tomioka, Y., Asamitsu, A., and Tokura, Y. 1999,
Phys.~Rev.~Lett. 82, 4316.

\bibitem{Moreo}
Moreo, A., Mayr, M., Feiguin, A., Yunoki, S., and Dagotto, E. 2000:
Phys.~Rev.~Lett. 84, 5568;
Mayr, M., Moreo, A., Verg\'{e}s, J.A., Arispe, J., Feiguin, A., and Dagotto, E. 2001,
Phys.~Rev.~Lett. 86, 135.

\bibitem{Nagaev}
Nagaev, E.L. 1996,
Physics-Uspekhi 39, 781; and references therein.

\bibitem{condmat2002}
Solovyev, I.V. 2003, e-print cond-mat/0212608.

\bibitem{COincommensurate}
Chen, C.H., and Cheong, S-W. 1996,
Phys.~Rev.~Lett. 76, 4042;
Chen, C.H., Mori.~S, and Cheong, S-W. 1999,
Phys.~Rev.~Lett. 83, 4792;
Jir\'{a}k, Z., Damai, F., Hervieu, M., Martin, C., Raveau, B., Andr\'{e}, G., and
Bour\'{e}e, F. 2000,
Phys.~Rev.~B 61, 1181;
Kajimoto, R., Yoshizawa, H., Tomioka, Y., and Tokura, Y. 2001,
Phys.~Rev.~B 63, 212407.

\bibitem{Arima}
Arima, T., Akahoshi, D., Oikawa, K., Kamiyama, T., Uchida, M., Matsui, Y., and
Tokura, Y. 2002,
Phys.~Rev.~B 66, 140408.

\bibitem{softening}
Hwang, H.Y., Dai, P., Cheong, S-W., Aeppli, G., Tennant, D.A., and Mook, H.A. 1998,
Phys.~Rev.~Lett. 80, 1316;
Fernandez-Baca, J.A., Dai, P., Hwang, H.Y., Kloc, C., and Cheong, S-W. 1998,
Phys.~Rev.~Lett. 80, 4012.

\bibitem{Motome}
Motome, Y., and Furukawa, N. 2002,
J.~Phys.~Soc.~Jpn. 71, 2002.

\bibitem{Kobayashi}
Kobayashi, K.-I., Kimura, T., Sawada, H., Terakura, K., and
Tokura, Y. 1998, Nature~(London)~395, 677;
Kobayashi, K.-I., Kimura, T., Tomioka, Y., Sawada, H., Terakura, K., and
Tokura, Y. 1999, Phys.~Rev.~B~59, 11159.

\bibitem{Tomioka}
Tomioka, Y., Okuda, T., Okimoto, Y., Kumai, R., Kobayashi, K.-I. and
Tokura, Y. 2000, Phys.~Rev.~B~61, 422.

\bibitem{PRB02}
Solovyev, I.V. 2002,
Phys.~Rev.~B 65, 144446.

\bibitem{KanamoriTerakura}
Kanamori, J. and Terakura, K. 2001,
J.~Phys.~Soc.~Jpn.~70, 1433.

\bibitem{KobayashiJMMM}
Kobayashi, K.-I., Okuda, T., Tomioka, Y., Kimura, T., and Tokura, Y. 2000,
J.~Magn.~Magn.~Matter.~218, 17.

\bibitem{Singh_Springer}
Singh, D.J. 2003, in
{\it Electronic Structure and Magnetism of Complex Materials},
D.J.~Singh and D.A.~Papaconstantopoulos (Eds.), Springer, Berlin.

\bibitem{Katsufuji}
Katsufuji, T., Hwang, H.Y., and Cheong, S-W. 2000,
Phys.~Rev.~Lett.~84, 1998.

\bibitem{Moritomo}
Moritomo, Y., Xu, Sh., Machida, A., Katsufuji, T., Nishibori, E., Takata, M.,
Sakata, M., and Cheong, S-W. 2001,
Phys.~Rev.~B~63, 144425.

\bibitem{Taguchi}
Taguchi, Y., Ohgushi, K., and Tokura, Y. 2002,
Phys.~Rev.~B~65, 115102.

\bibitem{Iikubo}
Iikubo, S., Yoshii, S., Kageyama, T., Oda, K., Kondo, Y., Murata, K., and Sato, M. 2001,
J.~Phys.~Soc.~Jpn.~70, 212.

\bibitem{Y2Mo2O7_basicExp}
Reimers, J.N., Greedan, J.E., and Sato, M. 1988,
J.~Solid~State~Chem.~72, 390;
Gardner, J.S., Gaulin, B.D., Lee, S.-H., Broholm, C., Raju, N.P., and
Greedan, J.E. 1999,
Phys.~Rev.~Lett.~83, 211.

\bibitem{Reimers}
Reimers, J.N. 1992, Phys.~Rev.~B~45, 7287.

\bibitem{Y2Mo2O7_recentExp}
Booth, C.H., Gardner, J.S., Kwei, G.H., Heffner, R.H., Bridges, F., and
Subramanian, M.A. 2000,
Phys.~Rev.~B~62, R755;
Keren, A., and Gardner, J.S. 2001,
Phys.~Rev.~Lett.~87, 177201.

\bibitem{Singh}
Singh, D.J. 1996,
J.~Appl.~Phys.~79, 4818.

\bibitem{Anisimov}
Anisimov, V.I., Korotin, M.A., Z\"{o}lfl, M., Pruschke, T., Le~Hur, K., and
Rice, T.M. 1998, Phys.~Rev.~Lett.~80, 5758.

\bibitem{PRB03.2}
Solovyev, I.V. 2003, Phys.~Rev.~B~97, 174406.

\bibitem{PRB96}
Solovyev, I., Hamada, N., and Terakura, K. 1996,
Phys.~Rev.~B~53, 7158.

\bibitem{Pickett}
Pickett, W.E. 1989,
Rev.~Mod.~Phys.~61, 433.


\end{thebibliography}
\end{document}